\documentclass[12pt,journal,onecolumn]{IEEEtran}
\usepackage[hang]{subfigure}
\usepackage{epsfig,rotating,setspace,latexsym,amsmath,epsf,amssymb,amsfonts,bm,theorem,epstopdf}
\usepackage{dsfont}
\usepackage{cite,authblk}
\usepackage{algorithm}
\usepackage[noend]{algpseudocode}
\usepackage{setspace}
\usepackage{color}
\usepackage{graphicx}
\usepackage{comment}
\usepackage{mathrsfs}
\ifCLASSOPTIONcompsoc
\usepackage[caption=false, font=normalsize, labelfont=sf, textfont=sf]{subfig}
\else
\usepackage[caption=false, font=footnotesize]{subfig}
\fi

\newtheorem{theorem}{Theorem}

\newtheorem{remark}{Remark}
\newtheorem{corollary}{Corollary}
\newtheorem{definition}{Definition}
\newtheorem{example}{Example}

\newenvironment{Proof}[1]{\medskip\par\noindent{\bf Proof:\,}\,#1}{{\mbox{\,$\blacksquare$}\par}}

\allowdisplaybreaks

\definecolor{green1}{rgb}{0.2,0.7,0.2}
\definecolor{brown}{rgb}{1,0.5,0.2}
\DeclareMathOperator*{\argmin}{\arg\!\min}
\DeclareMathOperator*{\argmax}{\arg\!\max}

\usepackage[colorinlistoftodos,textwidth=\marginparwidth]{todonotes}
\begin{document}

\title{Online and Offline Dynamic Influence Maximization Games Over Social Networks \thanks{This work is supported by the Air Force Office of Scientific Research FA9550-23-1-0107, the NSF CAREER Award EPCN-1944403, and the ARO MURI Grant AG285.}}
    \author{Melih Bastopcu \qquad S. Rasoul Etesami  \qquad Tamer Ba\c{s}ar\\
	\normalsize Coordinated Science Laboratory\\
	\normalsize University of Illinois Urbana-Champaign, Urbana, IL 61801\\
	\normalsize  \emph{(bastopcu,etesami1,basar1)@illinois.edu}}
\maketitle
\vspace{-2cm}
\begin{abstract}
In this work, we consider dynamic influence maximization games over social networks with multiple players (influencers). At the beginning of each campaign opportunity, individuals' opinion dynamics take i.i.d. realizations based on an arbitrary distribution. Upon observing the realizations, influencers allocate some of their budgets to affect their opinion dynamics. Then, individuals' opinion dynamics evolve according to the well-known De Groot model. At the end of the campaign time, the influencers collect their reward based on the final opinion dynamics. 
The goal of each influencer is to maximize their own reward subject to their limited total budget rate constraints. Thus, influencers need to carefully design their investment policies considering individuals' opinion dynamics and other influencers' investment strategies, leading to a dynamic game problem. We first consider the case of a single influencer who wants to maximize its utility subject to a total budget rate constraint. We study both \textit{offline} and \textit{online} versions of the problem where the opinion dynamics are either known or not known \textit{a priori}. In the singe-influencer case, we propose an online no-regret algorithm, meaning that as the number of campaign opportunities grows, the average utilities obtained by the offline and online solutions converge. Then, we consider the game formulation with multiple influencers in offline and online settings. For the offline setting, we show that the dynamic game admits a unique Nash equilibrium policy and provide a method to compute it. For the online setting and with two influencers, we show that if each influencer applies the same no-regret online algorithm proposed for the single-influencer maximization problem, they will converge to the set of $\epsilon$-Nash equilibrium policies where $\epsilon=\mathcal{O}(\frac{1}{\sqrt{K}})$ scales in average inversely with the number of campaign times $K$ considering the average utilities of the influencers. Moreover, we extend this result to any finite number of influencers under more strict requirements on the information structure. Finally, we provide numerical analysis to validate our results under various settings.                
\end{abstract}


\section{Introduction}
Advancements in communication technologies and the development of social media platforms in the past decade have enabled individuals to be connected more than ever. With these advancements, individuals can share their beliefs and alter their opinions as a result of interactions with other individuals and also with special entities called \textit{influencers}. For example, consider political campaigns for elections where political leaders want to shape public opinion to win elections. As another example, consider companies investing in online advertisements or influencers in social media platforms where companies or influencers want to affect public opinion so that individuals purchase their advertised products. In these examples, political parties or companies (in general, influencers) have limited budgets to allocate in order to affect individuals' opinions. Thus, the influencers need to carefully design their investment strategies to maximize their influence, perhaps without knowing the opinion realizations in the future, while considering the investment strategies of the other influencers and the interactions with other individuals within the social network. 

In this work, we consider a dynamic influence maximization game over social networks consisting of $n$ individuals and $m$ influencers who are interacting over $K$ campaign opportunities. At each campaign opportunity, a new set of opinions are realized based on a known distribution. Upon observing the realizations, the influencers invest their budgets in the individuals and shape their initial beliefs. After that, the individuals interact with each other based on the well-known De-Groot model \cite{degroot1974reaching}. At the end of each campaign period, influencers collect their corresponding rewards. In this influence maximization game, we consider the \textit{offline} (as well as \textit{online}) setting where the initial realizations of the opinion dynamics are (not) known by influencers \textit{a priori}. Our goal is to characterize the Nash equilibrium strategies for the influencers by considering both online and offline settings. 

\subsection{Related Works}
The influence maximization problem in social networks has been widely studied in the past literature. In \cite{ahmadinejad2015forming, Etesami2019, christia2021scalable}, the authors model the opinion dynamics in such a way that the internal and expressed beliefs of individuals can be different from each other. In \cite{ahmadinejad2015forming}, by altering the internal beliefs of individuals, the goal is to maximize the sum of external behaviors under a given budget constraint. In \cite{Etesami2019}, the authors investigate how the conformist and manipulative behaviors may affect public opinion. In \cite{kempe2003maximizing, gionis2013opinion, Chen2010}, the authors consider the problem of identifying a subset of the most influential individuals (also known as the seed-selection problem) that can maximize the overall influence in a social network. The competitive version of the influence maximization problem has also been studied in the past literature, which leads to game theoretic formulations \cite{bacsar1998dynamic}. More specifically, the competitive diffusion game models, where the players' goals are to choose a subset of the nodes at the beginning of the game so that these nodes can influence as many nodes as possible through the diffusion process, have been considered in \cite{alon2010note, etesami2016complexity}. Reference \cite{alon2010note} investigates the relationship between the network's diameter (i.e., the maximum distance between any pair of nodes) and the existence of pure Nash equilibria in a competitive diffusion game. In \cite{etesami2016complexity}, the authors study the complexity of existence of a pure Nash equilibrium in the competitive diffusion game over general networks and provide some necessary conditions for the existence of one such equilibrium. Reference \cite{grabisch2018strategic} considers a game between two competing influencers with opposing beliefs. Each influencer can initially affect a single node with different impacts, and then nodes' opinions evolve according to the De-Groot model. They showed that the game has a unique equilibrium in mixed strategies when the influencers have different impacts. Reference \cite{maehara2015budget} extends the budget allocation problem considered for a single advertiser \cite{alon2012optimizing} to a game-theoretic framework with multiple advertisers called budget allocation game with bipartite influence model.                 
    
The works that are most closely related to our work are \cite{varma2018marketing, varma2019marketing, christia2021scalable, Etesami2022}. In \cite{christia2021scalable}, the authors consider the static influence maximization game with multiple influencers. They show the existence of a unique Nash equilibrium under certain convexity assumptions on the influencers' objective functions and provide a method to efficiently compute the Nash equilibrium strategies. A similar problem has been considered in \cite{varma2018marketing} where the opinions evolve based on the De-Groot model, and the authors provide a closed-form expression for the Nash equilibrium solution in the case of two influencers. Reference \cite{varma2019marketing} extends the results in \cite{varma2018marketing} to a dynamic case where the authors show that their proposed policy leads to higher utilities for both influencers compared to applying the static Nash equilibrium solution repeatedly. Reference \cite{Etesami2022} considers the dynamic version of the influence maximization game. It provides a method to compute \textit{open-loop} equilibrium strategies where the influencers know individuals' initial opinions. Different from aforementioned earlier works, we study in this paper \textit{closed-loop} equilibrium strategies for dynamic influence maximization games by considering both the \textit{offline} case where the opinion realizations are known to the influencers and the \textit{online} case where they are unknown \textit{a priori}. 

\subsection{Contributions and Organization} 

In this work, our goal is to find the Nash equilibrium strategies for a dynamic influence maximization game with multiple budget-constrained influencers where the opinion dynamics take independent and identical realizations at each campaign opportunity. Our contributions and organization of the paper are as follows: 
\begin{itemize}
    \item In Section~\ref{Sect:single}, we consider the influence maximization problem for a single influencer. By using tools from convex optimization, we provide a closed-form solution for the offline setting in which the influencer knows the opinion realizations \textit{a priori}. Then, we consider the online setting where the influencer observes the opinion dynamics realizations at the beginning of each campaign opportunity. Leveraging some ideas from \cite{balseiro2020dual}, we develop an algorithm (Algorithm~\ref{alg: postcalc}) to obtain an online allocation policy with a regret bound of $\mathcal{O}(\frac{1}{\sqrt{K}})$, where $K$ is the number of campaign opportunities. As a result, the average utilities obtained by the online and offline policies will converge to the same values as $K$ grows.
    \item In Section~\ref{sect:offline_game}, we consider the influence maximization game with multiple influencers for the offline case, where we show that the offline game has a unique Nash equilibrium and the equilibrium policies can easily be computed by applying the best response dynamics with $m=2$-influencers. For $m>2$ influencers, if each influencer plays according to any no-regret algorithm, their average response converges to an $\epsilon$-Nash policy. 
    \item In Section~\ref{sect:online_game}, we consider the online version of the influence maximization game. Here, we show that under a certain information structure and when $m=2$, if each influencer follows Algorithm~\ref{alg: postcalc}, their strategies will converge to an $\epsilon$-Nash equilibrium of the offline game where $\epsilon=\mathcal{O}(\frac{1}{\sqrt{K}})$. Thus, the average utilities obtained from the offline and online settings converge to the same values as $K$ increases. Moreover, this result can be extended to the multiplayer ($m>2$) influence maximization game under a more stringent full-information structure. Finally, we propose an alternative algorithm that requires less information but still performs well compared to the full-information setting required for the multiplayer influence maximization game. 
    \item Finally, in Section~\ref{Sect:num_res}, we provide numerical experiments to verify our theoretical results and show the outperformance of our proposed algorithms under various scenarios. 
\end{itemize}

\begin{figure}[t]	\centerline{\includegraphics[width=0.75\columnwidth]{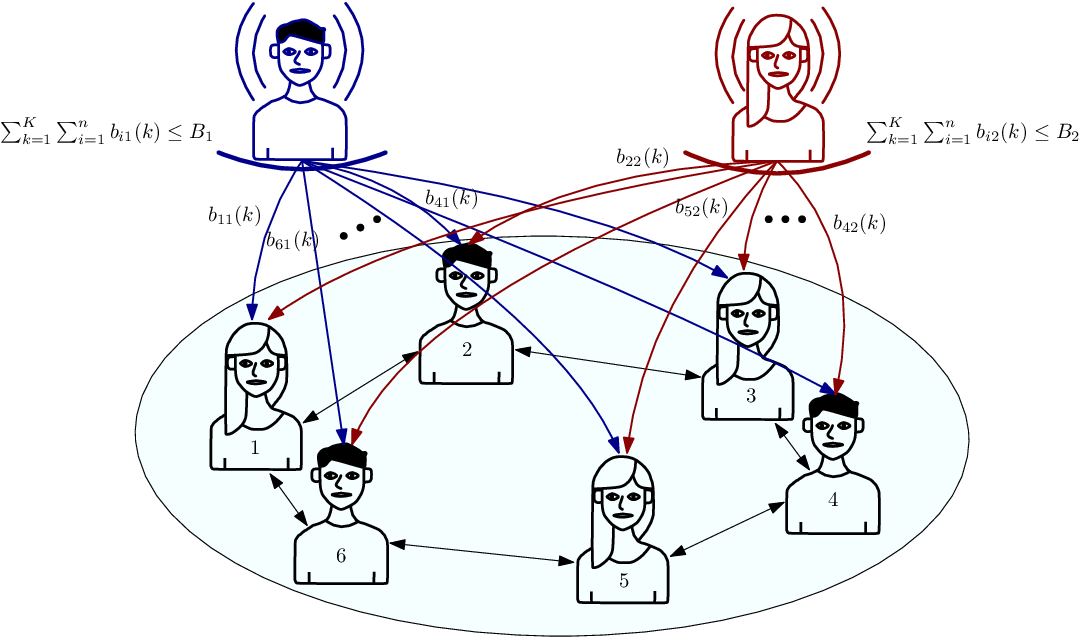}}
	\caption{A social network with $m=2$ influencers and $n=6$ individuals who are connected through a cycle. Here, by allocating some of their budgets $b_{ij}(k)$, influencers affect the opinion dynamics of the individuals.}
\label{Fig:System_model}
\vspace{-0.5cm}
\end{figure}

\section{System Model and Problem Formulation}\label{sect:model}

We consider a dynamic influence maximization game over social networks where there are $m$ influencers (players) and $n$ individuals interacting over a total time period $t\in[0,T]$ as shown in Fig.~\ref{Fig:System_model}. In this game, influencers have $K$ campaign times denoted by $t_0 = 0\leq t_1\leq \dots \leq t_K\leq t_{K+1}=T $. We define the inter-campaign times (or campaign duration) as $T_k = t_{k+1}-t_k$ for $k=1,\dots, K$, which take values from a set $\mathcal{T} = \{\tau_1, \dots, \tau_r\}$ based on a probability mass function (pmf) given by $\mathbb{P}_\mathcal{T}(\tau) =\{q_1, \dots, q_r\}$. We assume that each $T_k$ value is independent and identically distributed (i.i.d.). During these campaign times, influencer $j$ has the total budget $B_j>0$ to influence the opinion dynamics of the individuals. We denote by $B_j(k)$ the remaining budget of the $j$th influencer at campaign time $t_k$. 

At time $t$, we denote the opinion of the $i$th individual on the $j$th influencer by $x_{ij}(t)\in[0,1]$, and we have $\sum_{j=1}^{m}x_{ij}(t) = 1$ for all $i\in{1,\dots, n}$ and for all $t\in[0,T]$. We denote the matrix $x(t) \in [0,1]^{n\times m}$ as the opinions of all individuals over all influencers at time $t$. In this game, at the beginning of each campaign time $t_k$, the individuals' opinions $x(t_k)$ take an i.i.d. value from a set of matrices $\mathcal{X}=\{x^r|x^r\in [0,1]^{n\times m}, r= 1,\dots,d\}$ based on a probability distribution $\mathbb{P}_\mathcal{X}(x) = \{p_1,\dots,p_d\}$. In other words, at time $t_k$, we have $x(t_k) = x^r$ with probability $p_r$. This model can also be imagined as, at each campaign time, a new group of $n$ people drawn independently from the society where the group's prior opinion distribution is given by $\mathbb{P}_\mathcal{X}(x)$. This model is applicable in practical scenarios such as political campaigns to a portion of society or personal advertisements for individuals on commercial websites.

When the individuals' opinions and campaign duration take new realizations, the influencers observe $x(t_k)$ and campaign duration $T_k$ at time $t_k$. Then, the $j$th influencer invests $b_{ij}(t_k)\geq 0$ on the opinion of the $i$th individual with a constraint on the total remaining budget, $\sum_{i=1}^n b_{ij}(t_k)\leq B_j(k)$. We denote the investment vector of influencer $j$ at the campaign time $k$ by the row vector $b_{j}(t_k) =(b_{1j}(t_k),\ldots,b_{nj}(t_k))$. After the influencers make their investments, the individuals' opinions change proportional to these investments and according to the following model:
\begin{align}\label{eqn_opn_dyn_1}
    x_{ij}(t_k^+) = \phi (x_{ij}(t_k),b_{i,1:m}(t_k)) = \frac{x_{ij}(t_k) +b_{ij}(t_k)}{1+ \sum_{j=1}^m b_{ij}(t_k)},
\end{align}
where $t_k^+ = t_k+\epsilon$ with arbitrarily small $\epsilon>0$. Subsequently, the opinion dynamics of the individuals evolve based on the De-Groot model, which is given by $\dot{x}(t) =-Lx(t)$ for $t_{k-1}\leq t< t_k$, where $L$ is the weighted Laplacian matrix of the underlying social network among individuals, with appropriate dimensions. 

The rewards of the influencers for each campaign time $k$ are obtained based on the opinions of the individuals just before the next campaign time denoted by $t_{k+1}^- = t_{k+1}-\epsilon$ for arbitrarily small $\epsilon>0$. Let us define the matrix $A_k = e^{-L T_k}$ for $k = 1,\dots,K$. Then, the row vector $\rho_k = [\rho_{1k},\rho_{2k},\dots, \rho_{nk}]$ is given by $\rho_k = \mathds{1}^{\top} A_k$, where $\mathds{1}$ denotes a column vector of all {\it ones} with an appropriate dimension. The reward obtained by the $j$th influencer at time $k$ is given by 
\begin{align}\label{eqn:utility}
    u_j(x(t_k),\rho_{k}, b_{j}(t_k), b_{-j}(t_k)) \!=\!\! \sum_{i=1}^n \!\rho_{ik}  x_{ij}(t_k^+), 
\end{align}
where $b_{-j}(t_k)$ denotes the investment strategies of the other influencers at the $k$th campaign opportunity. 
The goal of influencer $j$ is to maximize its aggregate reward over all campaign times. Thus, influencer $j$ wants to solve the following optimization problem: 
\begin{align}\label{optimization_problem_1}
\max_{b_{ij}(t_k)} & \quad \sum_{k=1}^{K}u_j(x(t_k),\rho_{k}, b_{j}(t_k), b_{-j}(t_k)) \nonumber\\
\textrm{s.t.} & \quad \sum_{k=1}^{K}\sum_{i=1}^{n} b_{ij}(k)\leq B_j \nonumber\\
&\quad 0\leq b_{ij}(k)\leq \bar{b}_j, \quad \forall i,k, 
\end{align}
where each summand in the objective function of (\ref{optimization_problem_1}) is given by (\ref{eqn:utility}). Here, the first constraint captures the total investment budget of influencer $j$, and the last ones are the feasibility constraints. This defines a noncooperative game among the influencers, where the objective function and the strategy set of player $j$ are given by those in \eqref{optimization_problem_1}.  
\section{Optimal Investment Strategy for a Single Influencer}\label{Sect:single}

In order to understand the optimal investment strategies better, we focus in this section on the single-influencer setting and provide an optimal solution to the single-influencer maximization problem. We postpone the analysis for the game-theoretic setting with multiple influencers to subsequent sections. 

For the single influencer maximization problem, we consider both the offline and online settings. In the offline setting, the influencer knows the realizations of the opinion dynamics and the campaign times in advance, and thus the influencer can decide on its investment policy optimally. In the online setting, the influencer can only observe the opinion dynamics of the individuals, $\!x(t_k)$, and the duration of the campaign time $T_k$ at time $t_k$; it cannot observe the realizations of the future. Thus the influencer has to find an efficient investment strategy such that the total expected utility obtained in the online setting is close enough to the expected utility in the offline setting. As there is only one influencer in this section, we drop the influencer index. 

\subsection{The Optimal Investment Strategy in Offline Setting}\label{subsect:offline}
In the offline setting, all the realizations of $x(t_k)$ and $T_k$ are known at time $t=0$. Thus, the optimal investment strategy can be found by solving the optimization problem: 
\begin{align}\label{optimization_problem_single}
\max_{b_{i}(k)} & \quad \sum_{k=1}^{K} \sum_{i=1}^n \rho_{ik}  x_{i}(t_k^+) \nonumber\\
\textrm{s.t.} & \quad \sum_{k=1}^{K}\sum_{i=1}^{n} b_{i}(k)\leq B \nonumber\\
&\quad 0 \leq b_{i}(k)\leq\bar{b}, \quad \forall i,k,
\end{align}
where the objective function in (\ref{optimization_problem_single}) is the total utility obtained from the individuals with the constraint on the total investment budget. We note that from \eqref{eqn_opn_dyn_1}, the optimization problem in (\ref{optimization_problem_single}) is concave in terms of the budget variables $b_i(k)$, and thus the optimal solution can be obtained by analyzing the KKT conditions. For that, we define the Lagrangian function as follows: 
\begin{align}\label{lagrange}
    \!\mathcal{L} =& \!-\!\sum_{k=1}^{K} \sum_{i=1}^n \rho_{ik}  x_{i}(t_k^+)  +\theta \left( \sum_{k=1}^{K}\sum_{i=1}^{n} b_{i}(k)\!-\! B\!\right)\!-\!\sum_{k=1}^{K}\sum_{i=1}^{n} \nu_{ik} b_{i}(k)+\sum_{k=1}^{K}\sum_{i=1}^{n} \gamma_{ik} (b_{i}(k)-\bar{b}),\!
\end{align}
where $\theta \geq 0$, $\nu_{ik}\geq 0$, and $\gamma_{ik}\geq 0$ for all $i\in[n]$ and $k\in[K]$. Next, we write the KKT conditions as
\begin{align}\label{KKT1}
\!\!\frac{\partial \mathcal{L}}{\partial b_{i}(k)} \!=\!
\frac{\rho_{ik}(-1 +x_{i}(t_k))}{(1+ b_{i}(k))^2} +\theta - \nu_{ik}+\gamma_{ik} = 0,\hspace{1mm} \forall i, k. 
\end{align}
The complementary slackness conditions are given by 
\begin{align}
    \theta \left( \sum_{k=1}^{K}\sum_{i=1}^{n} b_{i}(k)- B\right) =& 0,\label{CS1}\\
    \nu_{ik} b_{i}(k)= & 0,\quad \forall i, k,\label{CS2}\\
    \gamma_{ik} (b_{i}(k)-\bar{b}) =& 0,\quad \forall i, k.\label{CS3}
\end{align}
As (\ref{optimization_problem_single}) is a concave optimization problem, the KKT and complementary slackness conditions in (\ref{KKT1})-(\ref{CS3}) provide necessary and sufficient conditions for the optimal solution. From (\ref{KKT1}), we have 
\begin{align}\label{eqn:temp}
    b_{i}(k) = \sqrt{\frac{\rho_{ik}(1 -x_{i}(t_k))}{\theta - \nu_{ik}+\gamma_{ik}}}-1.
\end{align}
From (\ref{CS2}), we have either $b_{i}(k)>0$ and $\nu_{ik} = 0 $, or $b_{i}(k)=0$ and $\nu_{ik} \geq 0 $. From (\ref{CS3}), we have either $b_{i}(k)<\bar{b}$ and $\gamma_{ik} = 0 $, or $b_{i}(k)=\bar{b}$ and $\gamma_{ik} \geq 0 $. Thus, we can rewrite (\ref{eqn:temp}) as
\begin{align}\label{eqn:soln_bik}
    b_{i}(k) =\min \left\{ \left(\sqrt{\frac{\rho_{ik}(1 -x_{i}(t_k))}{\theta }}-1\right)^+, \bar{b} \right\},
\end{align}
where $(\cdot)^+ = \max\{0,\cdot\}$. 



Since $x_{i}(t_k^+)$ is an increasing function of $b_{i}(k)$, in the optimal solution, the influencer should use its entire budget $B$, i.e., $\sum_{k=1}^{K}\sum_{i=1}^{n} b_{i}(k)= B$. In this case, due to (\ref{CS1}), we have $\theta \geq 0$. In order to find the optimal $\theta$, we first ignore the operators $(\cdot)^+$ and $\min\{\cdot\}$ in (\ref{eqn:soln_bik}) and find a $\theta$ that solves for $\sum_{i=1}^{n} b_i(k) = B$. After finding $\theta$, we substitute it back into (\ref{eqn:soln_bik}) without the operators $(\cdot)^+$ and $\min\{\cdot\}$. We first check if all $b_i(k) \leq \bar{b}$. If there are $b_i(k) > \bar{b}$, we set the highest $b_i(k) =\bar{b} $, and solve for the remaining $b_i(k)$ values again until all $b_i(k) \leq \bar{b}$. Then, we check the $b_i(k) \geq 0$ conditions. If all $b_i(k) \geq 0$, we obtain the optimal $\theta$, and thus $b_i(k)$ values. If we have $b_i(k)<0$, then we choose the individual with the smallest $\rho_{ik}(1 -x_{i}(t_k))$ and take $b_i(k) = 0$. For the remaining individuals, we repeat this process (i.e., by ignoring $(\cdot)^+$ and $\min\{\cdot\}$ operators) and solve $\sum_{i=1}^{n} b_i(k) = B$ for $\theta$ and determine the remaining $b_i(k)$ values. If there is still $b_i(k)<0$, we choose the next individuals with the smallest $\rho_{ik}(1 -x_{i}(t_k))$ value and choose $b_i(k) = 0$. We repeat this process until all $b_i(k)\geq 0$. We note that the above process generates the optimal budget investments due to the following reason: the highest $b_i(k)$ value has the highest $\rho_{ik}(1 -x_{i}(t_k))$ term as seen in (\ref{eqn:soln_bik}). Thus, if there exists a $b_i(k)$ term that exceeds $\bar{b}$ (when we ignore the $\min(.,\bar{b})$ operator), equating that $b_i(k) =\bar{b}$ and solving for the remaining $\sum_{k=1}^{K}\sum_{i=1}^{n} b_{i}(k)= B$ decreases $\theta$ in the next round. Thus, the terms that have been equated to $\bar{b}$ in the earlier rounds will still remain as $b_{i}(k) = \bar{b}$. Similar arguments can be also made for the negative $b_i(k)$ (after ignoring the $(.)^+$ operator). Thus, by ignoring $(.)^+$ and $\min(.,\bar{b})$ operators initially, and placing them back by following the order of $\rho_{ik}(1 -x_{i}(t_k))$ one by one help us to find the optimal $\theta$ value which gives us the overall optimal budget allocation policy.    


\subsection{Optimal Investment Strategy in Online Setting}\label{subsect:offline}

In this subsection, we consider the online setting where the influencer does not observe the future realizations of the opinion dynamics and campaign durations and has to design its policy based on observing the current opinion dynamics $x_{i}(t_k)$, the campaign duration $T_k$ (or, equivalently $\rho_{ik}$), and its remaining budget $B(k)$. For this problem and using similar ideas as in \cite{balseiro2020dual}, we propose Algorithm~\ref{alg: postcalc}, which updates the estimate of the dual variable $\theta$ (in the offline setting) based on the realizations of $x_{i}(t_k)$, $\rho_{ik}$, and the remaining budget $B(k)$. In the online setting, our main goal is to achieve a sub-linear regret bound over an arbitrary opinion and campaign time distribution $\mathbb{P}_{\mathcal{X}\times \mathcal{T}}$ such that the average expected utility obtained from the online algorithm converges to the average utility obtained from the offline solution as the time horizon $K$ grows. To that end, let us rewrite the utility function obtained from individual $i$ at campaign opportunity $k$ as follows:
\begin{align}\label{eqn:utility_single}
    u_{ij}(x_i(t_k), \rho_{k}, b(k)) =  \rho_{ik} \frac{x_{i}(t_k) +b_{i}(k)}{1+ b_{i}(k)},
\end{align}
with $ u_{ij}(x_i(t_k),\rho_{ik},0) = \rho_{ik}x_{i}(t_k)$. Then, we define a new adjusted utility function as: 
\begin{align}\label{eqn:utility_v2}
    \tilde{u}_{ij}(x_i(t_k),\rho_{k}, b_i(k)) = u_{ij}(x_i(t_k), \rho_{k},  b_i(k))  - u_{ij}(x_i(t_k),\rho_{k},0)= \frac{ \rho_{ik} b_{i}(k)(1\!-\!x_{i}(t_k))}{1+ b_{i}(k)}.
\end{align}
We note that since there is a constant difference between $\tilde{u}_{ij}(x_i(t_k), \rho_{k},b_i(k))$ and $u_{ij}(x_i(t_k), \rho_{k},$ $ b_i(k))$, the utility maximization problem remains the same. 

In the online setting, we use a budget allocation policy $Q$ to make an investment decision at time $t_k$, which depends on the history $H_{k-1} = \{x(t_{\ell}), \rho_{\ell}, b(\ell)\}_{\ell=1}^{k-1}$, the current realization of the individuals' opinions $x(t_k)$, and the campaign duration $T_k$. Thus, we have 
\begin{align}
    \hat{b}(k) = Q(x(t_{k}),\rho_{k}|H_{k-1} ).
\end{align}
By applying policy $Q$, the aggregate total utility obtained over all individuals is given by $\tilde{u}_j(x(t_k), \rho_{k}, \hat{b}(k))$ $=\sum_{k=1}^{K} \sum_{i=1}^{n} \tilde{u}_{ij}(x_i(t_k), \rho_{ik}, \hat{b}_{i}(k))$. Thus, the total expected utility over all possible realizations under the joint distribution of opinions and the campaign times $\mathbb{P}_{\mathcal{X}\times \mathcal{T}}$ equals
\begin{align}\label{eqn_online}
 \!\!\tilde{U}(Q|\mathbb{P}_{\mathcal{X}\times \mathcal{T}}) \!=\! \sum_{k=1}^{K}\sum_{i=1}^n \mathbb{E}_{\mathbb{P}_{\mathcal{X}\times \mathcal{T}}}\!\left[\tilde{u}_{ij}(x_i(t_k),\rho_{ik},\hat{b}_i(k))\right]\!.
 \end{align}

In the offline setting where all realizations of $\{x_i(t_k),$ $\rho_{ik}\}$ are known by the influencer, the overall utility is given by the solution of the following optimization problem: 
\begin{align}\label{eqn:primal_problem}
 \!\!\!\tilde{U}_{opt}(\{x(t_k),{\rho_k}\}_{k = 1}^{K}) =
\max_{b_{i}(k)} & \quad \sum_{k=1}^K \sum_{i=1}^n\tilde{u}_{ij}(x_i(t_k),\rho_{ik}, b_i(k))\nonumber\\
\textrm{s.t.} & \quad \sum_{k=1}^{K}\sum_{i=1}^{n} b_{i}(k)\leq B \nonumber\\
&\quad 0\leq b_{i}(k)\leq \bar{b}, \quad \forall i,k.
\end{align}

Then, the total expected offline utility is given by 
\begin{align}\label{eqn_offline}
     \tilde{U}_{opt} (\mathbb{P}_{\mathcal{X}\times\mathcal{T} }) = \mathbb{E}_{\mathbb{P}_{\mathcal{X}\times\mathcal{T}}}[ \tilde{U}_{opt}(\{x(t_k),\rho_k\}_{k = 1}^{K})].
\end{align}

We define the regret of following the policy $Q$ as
\begin{align}
    R(Q|\mathbb{P}_{\mathcal{X}\times \mathcal{T}}) = \tilde{U}_{opt} (\mathbb{P}_{\mathcal{X}\times \mathcal{T}}) - \tilde{U}(Q|\mathbb{P}_{\mathcal{X}\times \mathcal{T}}).
\end{align}

Moreover, we define the supremum of the regret over the family of feasible distributions $ \mathbb{P}_{\mathcal{X}\times \mathcal{T}}\in \mathcal{J}$ as
\begin{align}\label{eqn_sup_regret}
    R(Q|\mathcal{J}) =  \sup_{\mathbb{P}_{\mathcal{X}\times \mathcal{T}}\in \mathcal{J}} \{\tilde{U}_{opt} (\mathbb{P}_{\mathcal{X}\times \mathcal{T}}) - \tilde{U}(Q|\mathbb{P}_{\mathcal{X}\times \mathcal{T}})\}.
\end{align}

In Algorithm~\ref{alg: postcalc}, we denote the dual variable at the $k$th investment time as $\theta_k$. Then, after observing opinion dynamics of the individuals, $x_i(k)$, and the campaign duration $T_k$, which determines $\rho_{ik}$, the influencer makes its investment decision for a given $\theta_k$ as follows:
\begin{align}\label{optimization_problem_single_online}
\tilde{b}_{i}(k) = \argmax_{0\leq b_{i}(k)\leq \bar{b}}  \sum_{i=1}^n \frac{ \rho_{ik} b_{i}(k)(1-x_{i}(t_k))}{1+ b_{i}(k)}+\theta_k b_{i}(k).
\end{align}
By noting that (\ref{optimization_problem_single_online}) is a concave function with respect to $b_{i}(k)$, we can find the optimum $\tilde{b}_{i}(k)$ that maximizes (\ref{optimization_problem_single_online}) for a given $\theta_k$ as
\begin{align}\label{eqn:opt_online}
  \tilde{b}_{i}(k) = \min \left\{ \left( \sqrt{\frac{\rho_{ik}(1-x_i(t_k))}{\theta_k}}-1\right)^+, \bar{b}\right\}. 
\end{align}

We note that the optimal solution $\tilde{b}_{i}(k)$ in (\ref{eqn:opt_online}) in the online setting has similar expression with $b_i(k)$ given in (\ref{eqn:soln_bik}) for the offline problem where $\theta$ in (\ref{eqn:soln_bik}) is replaced with $\theta_k$ in (\ref{eqn:opt_online}). In the offline setting, since all $\{x_i(t_k), \rho_{ik}\}$ are known by the influencer, one can determine the optimal $\theta$. On the other hand, in the online setting, the influencer only observes the opinions and the campaign duration at the current stage and must determine its allocation strategy by deciding on the $\theta_k$ variable. Therefore, we can think of $\theta_k$ in the online problem as an estimate of $\theta$ in the offline problem. 

Next, we check the summation of $\tilde{b}_{i}(k)$ values in (\ref{eqn:opt_online}). If we have $\sum_{i=1}^n \tilde{b}_{i}(k)\leq B(k)$, where $B(k)$ denotes the remaining budget of the influencer at the investment time $k$, then the influencer uses $\tilde{b}_{i}(k)$ in (\ref{eqn:opt_online}) as its investment decision. However, if $\sum_{i=1}^n \tilde{b}_{i}(k)> B(k)$, then the influencer chooses $ \hat{b}_i(k) = 0$. Thus, we can summarize the influencer's investment decision in the online setting as 
\begin{align}\label{eqn:opt_onl_investment}
\hat{b}_i(k) = \begin{cases}
\tilde{b}_{i}(k), & \text{if $\sum_{i=1}^n \tilde{b}_{i}(k)\leq B(k)$, } \\
0, & \text{otherwise}.
\end{cases}
\end{align}
Based on the investment decision at time $t_k$, we update the remaining budget of the influencer at the $(k+1)$th campaign time as
\begin{align}\label{eqn:remaining_budget}
    B(k+1) = B(k) - \sum_{i=1}^n \hat{b}_i(k). 
\end{align}

\begin{algorithm}[t]
\begin{small}
\caption{Online Dual Mirror Descent Type Algorithm to Solve (\ref{optimization_problem_single}) }\label{alg: postcalc}
\begin{algorithmic}[1]
\State{\textbf{Parameters:} $x\in \mathcal{X}$, $ \tau\in \mathcal{T}$ with $\mathbb{P}_{\mathcal{X}\times \mathcal{T}}$, campaign times $k\in K$, Laplacian matrix $L$, average budget $\alpha$, step size $\eta$, initial dual variable $\theta_0$, remaining budget $B(k)$ with $ B(1) = B = \alpha K$  }
\State{\textbf{for} $ k=1,\dots, K$}
\State{\textbf{\quad Receive:} $(x(k),\tau_k)\sim\mathbb{P}_{\mathcal{X}\times \mathcal{T}}$, i.e., $x(k) = x^r$ with prob. $p_r$, and \newline  $ T_k = \tau_\ell$ with prob. $q_\ell$ }
\State{\textbf{\quad Compute:} $\tilde{b}_i(k)$ from (\ref{eqn:opt_online})}.
\State{\textbf{\quad Find:} $\hat{b}_i(k)$ from (\ref{eqn:opt_onl_investment})}
\State{\textbf{\quad Update:} $B(k+1) = B(k) - \sum_{i=1}^n \hat{b}_i(k)$ as in (\ref{eqn:remaining_budget})}
\State{\textbf{\quad Find a stochastic sub-gradient of $D(\theta_k)$ in (\ref{dual_function2}):}}
\begin{align*}
    \tilde{g}(k) = -\sum_{i=1}^n \tilde{b}_i(k)+\alpha \text{ as in (\ref{eqn:g(t_k)})} 
\end{align*}
\State{\textbf{\quad Update the dual variable by mirror descent:}}
\begin{align*}
    \theta_{k+1} = \argmin_{\theta\geq 0} \tilde{g}(k)\theta + \frac{1}{\eta} V_h(\theta,\theta_{k} ) \text{ as in (\ref{eqn:mirror_descent})}
\end{align*}
\State{\textbf{end}} 
\end{algorithmic}
\end{small}
\end{algorithm}

In order to find the dual variable for the next investment time, we write the dual function for the total expected utility optimization problem in (\ref{eqn_offline}) as
\begin{align}\label{dual_function}
\bar{D}(\theta) =\mathbb{E}_{\mathbb{P}_{\mathcal{X}\times\mathcal{T}}}\! &\left[ \max_{0 \leq b_{i}(k)\leq \bar{b}} \sum_{k=1}^K \left(\sum_{i=1}^n \frac{\rho_{ik}b_i(k)(1-x_i(t_k))}{1+b_i(k)} -\theta\left(\sum_{i=1}^n b_i(k)-\frac{B}{K} \right)\right)\right].
\end{align}

We note that since the values $\{x_i(t_k),\rho_{ik}\}$ are realized i.i.d. over different campaign times, we can rewrite (\ref{dual_function}) as 
\begin{align}\label{dual_function2}
\bar{D}(\theta) = K\mathbb{E}_{\mathbb{P}_{\mathcal{X}\times\mathcal{T}}} &\left[\max_{0 \leq b_{i}(k)\leq \bar{b}}   \sum_{i=1}^n \frac{\rho_{ik}b_i(k)(1\!-\!x_i(t_k))}{1+b_i(k)}-\theta\left(\sum_{i=1}^n b_i(k) \right)\right] +\theta B.
\end{align}
If we drop the constant term $K$ (as it does not affect the optimization problem), we can define
\begin{align}\label{dual_function_final}
D(\theta) = \mathbb{E}_{\mathbb{P}_{\mathcal{X}\times \mathcal{T}}}& \left[\max_{0 \leq b_{i}(k)\leq \bar{b}}  \sum_{i=1}^n \frac{\rho_{ik}b_i(k)(1-x_i(t_k))}{1+b_i(k)}-\theta\left(\sum_{i=1}^n b_i(k) \right)\right]+\frac{\theta B}{K}.
\end{align}
Since the optimization problem in (\ref{dual_function_final}) is separable over the variables $b_{i}(k)$, we can define the dual function for the $i$th individual as 
\begin{align}\label{dual_function_final_i}
D_i(\theta) \!=\! \mathbb{E}_{\mathbb{P}_{\mathcal{X}\times \mathcal{T}}} \left[\max_{0 \leq b_{i}(k)\leq \bar{b}}   \frac{\rho_{ik}b_i(k)(1-x_i(t_k))}{1+b_i(k)}-\theta b_i(k) \right]+\frac{\theta B}{nK},
\end{align}
where we have $D(\theta) = \sum_{i=1}^{n}D_i(\theta)$. Therefore, the offline dual problem for the total expected utility optimization becomes $\min_{\theta\geq 0} D(\theta)$. We note that (\ref{dual_function_final}) can be thought of as the dual problem defined over an average expected utility obtained from a single campaign time with average budget $\alpha = \frac{B}{K}$. 
By evaluating the subgradient of the offline dual problem at $\theta = \theta_k$, we have
\begin{align}\label{eqn:g(t_k)}
    \tilde{g}(k) = \sum_{i=1}^{n} \left.\frac{\partial D_i(\theta)}{\partial \theta} \right|_{\theta = \theta_k}\!\!\! =  -\sum_{i=1}^n \tilde{b}_i(k)+\alpha.
\end{align}
We use the function $\tilde{g}(k)$ to update the variable $\theta_k$ by applying an online mirror descent with step size $\eta$. For that, we define the Bregman divergence as $V_h(x,y) = h(x)- h(y) -\langle\nabla h(y), x - y\rangle$, where $h(\cdot)$ is a convex regularizer function and $\langle \cdot, \cdot\rangle$ denotes the inner product. Then, we update the dual variable $\theta_k$ as follows:
\begin{align}\label{eqn:mirror_descent}
    \theta_{k+1} = \argmin_{\theta\geq 0} \tilde{g}(k)\theta + \frac{1}{\eta} V_h(\theta,\theta_{k} ).
\end{align}

\begin{example}
By choosing the entropy function $h(x) = -\sum_{i=1}^n x_i\log(x_i)$ as a regularizer in the dual variable update, the mirror descent step in (\ref{eqn:mirror_descent}) becomes $\theta_{k+1} = \theta_{k} e^{-\eta \tilde{g}(t_k)}$. Therefore, for a given $\theta_k$, if the influencer spends more than its average budget $\alpha = \frac{B}{K}$ at time $k$, i.e., $\sum_{i=1}^n \tilde{b}_i(k)>\alpha$, then $\tilde{g}(k)$ in (\ref{eqn:mirror_descent}) becomes negative. As a result, in the mirror descent step we increase $\theta_{k+1}$ for the next investment time, which decreases the budget usage, i.e., $\tilde{b}_i(k+1)$ in (\ref{eqn:opt_online}). On the other hand, if the influencer uses less than its average budget at time $k$, i.e., $\sum_{i=1}^n \tilde{b}_i(k)<\alpha$, the mirror descent step  decreases $\theta_{k+1}$, which increases $\tilde{b}_i(k+1)$ for the next investment opportunity. Thus, the algorithm encourages the influencer to use its average resource in each step to ensure that it does not run out of budget quickly and also spend its entire budget at the end of $K$ investment opportunities.
\end{example}

\begin{theorem}\label{Thm_1}
Let $\eta \leq \frac{\sigma_2}{\bar{b}}$ and $ \theta_0 \leq \theta^{max}$. Then, for any $K\geq 1$, the regret of Algorithm \ref{alg: postcalc} satisfies
\begin{align}\label{eqn_thm_total_regret}
    R(Q|\mathcal{J})\leq& \frac{2 (n\bar{b}^2 +\frac{\alpha^2}{n})}{\sigma_1}\eta K + \frac{nV_h(0, \theta_0)}{\eta}+\frac{\bar{u}(\dot{h}(\theta^{max}) -  \dot{h}(\theta_{0}))}{\eta\alpha} + \frac{\bar{u}\bar{b}}{\alpha}.
\end{align}
In particular, by choosing $h(\theta) =\theta\log(\theta), \theta_0 = e^{-1}$, the regret $R(Q|\mathcal{J})$ in (\ref{eqn_thm_total_regret}) is bounded by 
\begin{align}\label{eqn_thm_total_regret_2}
  &R(Q|\mathcal{J})\leq \frac{\bar{u}\bar{b}}{\alpha} +2\sqrt{K} \sqrt{\frac{2}{\sigma_1}\left(n\bar{b}^2+\frac{\alpha^2}{n}\right)\left(ne^{-1} + \frac{\bar{u}(\log (\frac{\bar{u}}{\alpha}+1)+1)}{\alpha}\right)  }.
\end{align}
\end{theorem}

\begin{Proof}
The proof follows the same steps as in \cite{balseiro2020dual}. More specifically, it was shown in \cite{balseiro2020dual} for a different online resource allocation problem and under certain assumptions that one can obtain a sublinear regret bound as in the above theorem.

We first verify that all the assumptions in \cite{balseiro2020dual} indeed hold for the online influence maximization problem. Since $b_i(k)\in[0, \bar{b}]$ for all $i$ and $k$, the constraint set is bounded and convex. Also, each $b_i(k)$ can take a value of 0. Thus, Assumption~1 in \cite{balseiro2020dual} is satisfied.\footnote{Without loss of generality, we may assume that the opinions and campaign durations have finite supports, i.e.,  $|\mathcal{X}|\times |\mathcal{T}|$ is finite, where $|\cdot|$ denotes the cardinality of a set.} Moreover, we find an upper bound on $\sum_{i=1}^n\!\!\tilde{u}_{ij}(x_i(t_k), \rho_{ik}, b_i(k))$ as
\begin{align}
 \sum_{i=1}^n\!\tilde{u}_{ij}(x_i(t_k) ,\rho_{ik},b_i(k))\! &= \!\sum_{i=1}^n \frac{\rho_{ik}b_i(k)(1\!-\!x_i(t_k))}{1+b_i(k)} \leq \bar{u},
\end{align}
where $\bar{u} = \sum_{i=1}^n \rho_{ik}<\infty $, and the inequality follows by noticing $0\leq \frac{b_i(k)}{1+b_i(k)}\leq 1$, and $0\leq x_i(t_k)\leq 1$. Finally, there exists a $\bar{b}$ such that $||b(k)||_{\infty} \leq \bar{b}$ as we choose $b(k)$ from a compact space $[0, \bar{b}]$. Thus, Assumption~2 in \cite{balseiro2020dual} is satisfied. We assume that the average initial budget denoted by $\alpha$ is finite, i.e., $\alpha = \frac{B}{K}<\infty$. Finally, we take the reference function $h(\cdot)$ to be separable coordinate-wise, i.e., $h(x) = \sum_{i=1}^{n} h_i(x_i)$, where $h_i(x_i)$ is a convex function with respect to $x_i$, $\sigma_1$-strongly convex in $\ell_1$-norm in the set $\mathcal{D}$, i.e., $h(x_1)\geq h(x_2) + \langle\nabla h(x_2), x_1-x_2\rangle + \frac{\sigma_1}{2} \| x_1-x_2\|_1^2$ for all $x_1, x_2\in \mathcal{D}$, and $\sigma_2$-strongly convex in $\ell_2$-norm in the set $\mathcal{D}$. Thus, Assumptions~1-4 in \cite{balseiro2020dual} are satisfied. 


Next, let us define $\theta^{max} = \frac{\bar{u}}{\alpha} +1$. Using \cite[Lemma~2]{balseiro2020dual} and $\theta^+ = \argmin_{\hat{\theta}\geq 0} \tilde{g}(k), \hat{\theta} + \frac{1}{\eta} V_h(\hat{\theta},\theta_{k} )$ with $\tilde{g}(k)$ given in (\ref{eqn:g(t_k)}), for any $\theta \leq \theta^{max}$ and step size $\eta \leq \frac{\sigma_2}{\bar{b}}$, we have $\theta^+ \leq \theta^{max}$. Let us define the stopping time of the algorithm $\tau_Q$ to be the first time before $K$ such that 
\begin{align}
    \sum_{k=1}^{\tau_Q} \sum_{i=1}^{n} b_i(k) +\bar{b}\geq B. 
\end{align}
In other words, $\tau_Q$ is the first (random) time the influencer exhausts more than $B-\bar{b}$ amount of its budget. With a proper choice of step size $\eta\leq \frac{\sigma_2}{\bar{b}}$ and $\theta_0\leq\theta^{max}$, we know that $\theta_k\leq \theta^{max}$ for all $k\leq K$. Further, due to \cite[Proposition~2]{balseiro2020dual}, we have 
\begin{align}\label{eqn_Lemma_2}
    K-\tau_Q \leq \frac{\dot{h}(\theta^{max}) -  \dot{h}(\theta_{0})}{\eta\alpha}  + \frac{\bar{b}}{\alpha},  
\end{align}
which establishes an upper bound on the remaining time where the influencer will be close to running out of its initial budget.

We complete the proof by developing an upper bound on the primal-dual gap until the stopping time $\tau_Q$. To that end, let $\gamma_k$ be the random variable that determines the individuals' initial opinions at time $t_k$ and the campaign duration $T_k$. Then, the Lagrangian dual variable at stage $k$, $\theta_k$, depends on the set $\xi_k=\{\gamma_0,\dots, \gamma_k\}$. With Algorithm~\ref{alg: postcalc}, for a given proper step size $\eta$, due to \cite[Proposition~3]{balseiro2020dual}, we have
\begin{align} \label{Eqn:prop_3}
\mathbb{E}_{\mathbb{P}_{\mathcal{X}\times\mathcal{T}}} \left[ \tau_Q D_i( \bar{\theta}_{\tau_Q}) - \sum_{k=1}^{\tau_Q} \tilde{u}_i(x_i(t_k), \rho_{ik}, b_i(k)) \right]\leq \frac{2(\bar{b}^2+\frac{\alpha^2}{n^2})}{\sigma_1} \eta \mathbb{E}_{\mathbb{P}_{\mathcal{X}\times\mathcal{T}}}[\tau_Q] + \frac{V_h(0, \mu_0)}{\eta},
\end{align}
where $D_i(\cdot)$ is the individual's dual function given in (\ref{dual_function_final}) and $\bar{\theta}_{\tau_Q} = \frac{\sum_{k=1}^{\tau_Q} \theta_k}{\tau_Q}$. It is worth noting that different from \cite{balseiro2020dual}, here we develop an upper bound on the expected regret for each individual until $\tau_Q$ for a given total average utility $\frac{B}{nK} = \frac{\alpha}{n}$, which brings the additional $n^{-2}$ term in (\ref{Eqn:prop_3}). By combining (\ref{eqn_Lemma_2}) and (\ref{Eqn:prop_3}), we obtain the regret bound given in \eqref{eqn_thm_total_regret}.


Finally, when $h(\theta)=\theta \log (\theta)$, since the regret $R(Q|\mathcal{J})$ is bounded by (\ref{eqn_thm_total_regret}) for every $\eta$, the minimum of such upper bound is obtained by choosing $\eta = \sqrt{\frac{B}{A}}$, where $A =\frac{2K}{\sigma_1}\left(n\bar{b}^2+\frac{\alpha^2}{n}\right)$ and $B =ne^{-1} + \frac{\bar{u}(\log (\frac{\bar{u}}{\alpha}+1)+1)}{\alpha}$. Substituting these expressions into \eqref{eqn_thm_total_regret} completes the proof.  
\end{Proof}

\begin{remark}
From (\ref{eqn_thm_total_regret_2}) one can see that as the total campaign opportunities $K$ increases, the average regret $\frac{R(Q|\mathcal{J})}{K}$ converges to $0$ with a rate of $\mathcal{O}(\frac{1}{\sqrt{K}})$.
\end{remark}

\section{Nash Equilibrium Strategies for Offline Multiplayer Influence Maximization Game}\label{sect:offline_game}

In this section, we consider the offline influence maximization game with $m\geq 2$ influencers, where each influencer wants to maximize its own utility function leading to a constant sum game between the influencers. In this section, we consider the offline setting where the influencers know the opinion dynamics of individuals and the campaign durations. Each influencer also knows the initial budgets of the other players. More specifically, we denote the information structure of each influencer in the offline setting by $I_j$, which is given by 
\begin{align}\nonumber
I_j=\{\{x(t_k), T_k\}_{k\in[K]}, \{B_j\}_{j\in[m]}\}.
\end{align}

First, we note that as the opinions at the beginning of each stage are not affected by the previous stages, and in the offline setting these realizations are known ahead of time, the dynamic game with $n$ individuals over $K$ campaign opportunities is equivalent to a static game with $m$ influencers (players) and $nK$ individuals over a single campaign time. For this static game, influencer $j$ has a total budget of $B_j$ as before, which needs to be spent over a single campaign opportunity over $nK$ individuals. Since the realizations of $T_k$'s are known, $\rho_{ik}$ values can be found offline at the beginning of the game. By using this line of approach, in the following theorem, we show that the dynamic game has a unique Nash equilibrium. 

\begin{theorem}\label{Thm_offline}
    The offline influence maximization game with $m\ge 2$ influencers and information structure $\{I_j, j\in [m]\}$ admits a unique Nash equilibrium.    
\end{theorem}
\begin{Proof}
We first note that the dynamic game with $n$ individuals for $K$ campaign duration in the offline setting is equivalent to the offline static game with $nK$ individuals. More specifically, in the equivalent static game, the opinion dynamics are realized from the set of matrices
\begin{align}\nonumber
\bar{\mathcal{X}}=\Big\{[x(t_1)^\top, x(t_2)^\top,\dots,x(t_K)^\top]^\top \in[0,1]^{nK\times m}:\ \ \mathbb{P}\{x(t_k)=x^r\}=p_r, \forall k\in [K], r\in [d]\Big\}.
\end{align}
Our goal is to show that the equivalent offline static game admits a unique Nash equilibrium. To that end, we show that the total utility function of influencer $j$ is a concave function with respect to its own strategy $b_{ij}(k)$. The second derivative of the total utility function is given by 
\begin{align}
 &\frac{\partial^2 \sum_{k=1}^{K}\sum_{i=1}^{n} u_j(x(t_k),\rho_{k}, b_{j}(k), b_{-j}(k))}{\partial b_{ij}(k)^2} = - \frac{2 \rho_{ik}(1-x_{ij}(t_k)+\sum_{j'\neq j} b_{ij'}(k))}{(1+\sum_{j=1}^{m}b_{ij}(k))^3}<0, \label{hessian_1} \\
 &\frac{\partial^2 \sum_{k=1}^{K}\sum_{i=1}^{n} u_j(x(t_k),\rho_{k}, b_{j}(k), b_{-j}(k))}{\partial b_{ij}(k) \partial b_{i'j}(\ell)} =0,\label{hessian_2}
\end{align}
for all $i'\neq i$ and $\ell \neq k$. Thus, the Hessian of the total utility function of the $j$th influencer, $\sum_{k=1}^{K}\sum_{i=1}^{n} u_j(x(t_k),\rho_{k}, b_{j}(k), b_{-j}(k))$, is a matrix consisting of negative diagonal elements given in (\ref{hessian_1}), and zero elements in the non-diagonal positions due to (\ref{hessian_2}). Therefore, since the Hessian is a negative definite matrix, we conclude that the total utility function of influencer $j$ is a jointly concave function of $b_{ij}(k)$ for all $i\in[n]$ and $k\in[K]$. Next, we consider the weighted sum of total utility functions over all influencers with the uniform weights $\frac{1}{Kmn}$ defined by 
\begin{align}\label{eqn_sum_fnc}
    \sigma(b_{1},\dots, b_{m}) = \frac{1}{Kmn}\sum_{j=1}^{m}\sum_{k=1}^{K}\sum_{i=1}^{n} u_j(x(t_k),\rho_{k}, b_{j}(k), b_{-j}(k)) = \frac{1}{Kmn}\sum_{k=1}^{K}\sum_{i=1}^{n}\rho_{ik}.  
\end{align}
Thus, $\sigma(b_{1},\dots, b_{m})$ in (\ref{eqn_sum_fnc}) is a constant, and hence a concave function of $\{b_j\}_{j\in[m]}$. Next, we show that $\sum_{k=1}^{K}\sum_{i=1}^{n} u_j(x(t_k),\rho_{k}, b_{j}(k), b_{-j}(k))$ is a convex function of $b_{i,-j}$. For that, we have   
\begin{align}\label{eqn:hessian_der}
 \frac{\partial^2 \sum_{k=1}^{K}\sum_{i=1}^{n} u_j(x(t_k),\rho_{k}, b_{j}(k), b_{-j}(k))}{\partial b_{i\ell}(k)\partial b_{i\ell'}(k)}= \frac{2\rho_{ik}(x_{ij}(t_k)+b_{ij}(k) )}{(1+\sum_{j=1}^{m}b_{ij}(k))^3},
\end{align}
for all $\ell\neq j$ and $\ell'\neq j$. Then, we find 
\begin{align}
 \frac{\partial^2 \sum_{k=1}^{K}\sum_{i=1}^{n} u_j(x(t_k),\rho_{k}, b_{j}(k), b_{-j}(k))}{\partial b_{i\ell}(k)\partial b_{i'\ell'}(k')}= 0, \quad \forall\ell'\neq j, 
\end{align}
for all $i'\neq i$ or for all $k'\neq k$. Next, we construct the Hessian matrix of $\sum_{k=1}^{K}\sum_{i=1}^{n} u_j(x(t_k),$ $\rho_{k}, b_{j}(k), b_{-j}(k))$ with respect to $b_{i,-j}(k)$, which is denoted by $H_j\in \mathbb{R}^{(m-1) n K \times (m-1) n K}$. $H_j$ consisting of the matrices $\tilde{H_j}(\{k,i\},\{ k',i'\})\in \mathbb{R}^{(m-1) \times (m-1)}$ has the following structure: 
\begin{align}\label{eqn:H_j}\small
    \begin{bmatrix} 
    \!\!\tilde{H_j}(\{1,1\},\{ 1,1\})\!\!& \!\!\tilde{H_j}(\{1,1\},\{1,2\})\!\! & \!\!\dots \!\! &\!\! \tilde{H_j}(\{1,1\},\{ 1,n\})\!\!&\!\! \tilde{H_j}(\{1,1\},\{ 2,1\})\!\! & \dots &\!\! \tilde{H_j}(\{1,1\},\{ K,n\})\!\!\\
    \!\!\tilde{H_j}(\{1,2\},\{ 1,1\})\!\!& \!\!\tilde{H_j}(\{1,2\},\{ 1,2\})\!\! &\!\! \dots \!\! & \!\!\tilde{H_j}(\{1,2\},\{ 1,n\})\!\!& \!\!\tilde{H_j}(\{1,2\},\{ 2,1\})\!\! &\!\! \dots \!\!&\!\! \tilde{H_j}(\{1,2\},\{ K,n\})\!\!\\
    \vdots& \vdots & \ddots  & \vdots& \vdots & \dots & \vdots \\
    \!\!\tilde{H_j}(\{K,n\},\{ 1,1\})\!\!& \!\!\tilde{H_j}(\{K,n\},\{ 1,2\}) \!\!& \!\!\dots  \!\!&\!\! \tilde{H_j}(\{K,n\},\{ 1,n\})\!\!& \!\!\tilde{H_j}(\{K,n\},\{ 2,1\})\!\! &\!\! \dots \!\!&\!\! \tilde{H_j}(\{K,n\},\{ K,n\})\!\!
    \end{bmatrix}.
\end{align}
Here, $\tilde{H_j}(\{k,i\},\{ k',i'\}) = \boldsymbol{0}$ for all $\{k,i\}\neq \{ k',i'\} $, and $\tilde{H_j}(\{k,i\},\{ k',i'\})$ has the following form:
\begin{align} \label{eqn:H_tilde}
    \tilde{H_j}(\{k,i\},\{k,i\}) = \begin{bmatrix} 
    h & h & \dots & h \\
    \vdots &  \vdots &\ddots  &  \vdots\\
    h&  h& \dots     & h 
    \end{bmatrix},
\end{align}
where $h$ is given in (\ref{eqn:hessian_der}). We note that $ \tilde{H_j}(\{k,i\},\{k,i\})$ in (\ref{eqn:H_tilde}) has an eigenvalue of $(m-1)h = \frac{2(m-1)\rho_{ik}(x_{ij}(t_k)+b_{ij}(k) )}{(1+\sum_{j=1}^{m}b_{ij}(k))^3}$, and the others are equal to $0$. Then, $H_j$ in (\ref{eqn:H_j}) has $n\times K$ eigenvalues, one of which being $(m-1)h = \frac{2(m-1)\rho_{ik}(x_{ij}(t_k)+b_{ij}(k) )}{(1+\sum_{j=1}^{m}b_{ij}(k))^3}$, and the remaining $(m-2)\times n\times K$ eigenvalues equal to $0$. Thus, $H_j$ in (\ref{eqn:H_j}) is a positive semi-definite matrix, and hence $\sum_{k=1}^{K}\sum_{i=1}^{n} u_j(x(t_k),$ $\rho_{k}, b_{j}(k), b_{-j}(k))$ is a convex function with respect to $b_{i,-j}(k)$.    

Let us define the pseudo-gradient of $\sigma(b_{1},\dots, b_{m})$ as
\begin{align}
    v(b_{1},\dots, b_{m}) = \left(\left\{\nabla_{b_j}\sum_{k=1}^{K}\sum_{i=1}^{n} u_j(x(t_k),\rho_{k}, b_{j}(k), b_{-j}(k))\right\}_{j\in[m]}\right)^\top. \nonumber
\end{align}
We define $V(b_{1},\dots, b_{m})\in \mathbb{R}^{mKn \times mKn}$ as the Jacobian of $ v(b_{1},\dots, b_{m})$ with respect to $(b_{1},\dots, b_{m})$. Since $\sum_{k=1}^{K}\sum_{i=1}^{n} u_j(x(t_k),$ $\rho_{k}, b_{j}(k), b_{-j}(k))$ is a strictly concave function in terms of variables $b_{ij}(k)$, and each $\sum_{k=1}^{K}\sum_{i=1}^{n} u_j(x(t_k),$ $\rho_{k}, b_{j}(k), b_{-j}(k))$ is a convex function in $b_{i-j}(k)$, and there exist weights $\frac{1}{Kmn}$ such that $\sigma(b_{1},\dots, b_{m})$ in (\ref{eqn_sum_fnc}) is a constant (and hence concave in $(b_{ij}(k),b_{i,-j}(k)))$, by using \cite{Goodman1965NoteOE}, we conclude that the matrix $V(b_{1},\dots, b_{m})+V^\top(b_{1},\dots, b_{m})$ is a negative definite matrix. This, in view of \cite[Theorem~2]{rosen1965existence}, shows that the offline game admits a unique Nash equilibrium point.  
\end{Proof}

When there are only $m=2$ influencers in the offline influence maximization game, one can compute the Nash equilibrium strategies efficiently, as noted in the following corollary, whose proof follows directly from \cite{gilboa1991social} by applying the best response dynamics to strictly diagonal two-player concave games.    
\begin{corollary}\label{corr1}
    For the 2-player offline influencer game, starting with any arbitrary initial policies, the influencers can reach their Nash equilibrium by implementing the best response dynamics.  
    \end{corollary}

 In fact, for the case of $m=2$ influencers, one can characterize the best response dynamics in a closed-form. For a given strategy of the other influencer $-j$, we solve the optimization problem in (\ref{optimization_problem_1}), and find the best response dynamics as follows: \begin{align} \label{eqn:best_response}
    b_{ij}(k)^* = BR_j(x_{ij}(k),b_{i-j}(k)) = \min \left( \left( \sqrt{\frac{\rho_{ik}(x_{i-j}(k)+b_{i-j}(k))}{\theta_j}} \!-\!1-b_{i-j}(k) \right)^+\!\!\!, \bar{b}_j\!\right),
\end{align}
where $\theta_j\geq 0$ is the Lagrange variable that can be found by satisfying the total budget constraint $\sum_{k =1}^{K}\sum_{i =1}^{n}b_{ij}(k) = B_j$, and $BR_j(x_{ij}(k),b_{i-j}(k))$ denotes the best response of influencer $j$ with respect to $x_{ij}(k)$ and $b_{i-j}(k)$ for all $k\in [K], i\in[n]$. The Nash equilibrium is obtained when the following two conditions are satisfied:
\begin{align}\label{eqn_Nash_response}
  b_{ij}^*(k) = BR_j(x_{ij}(k),b_{i-j}(k)^*),\  j=1,2.
\end{align}
Therefore, if the two influencers successfully apply their best response dynamics in (\ref{eqn_Nash_response}), their responses will converge to their Nash equilibrium strategies. On the other hand, the best response dynamics for $m\geq 2$ offline influencer game may diverge in general. For that reason, we provide the following corollary, which provides a method to find \emph{$\epsilon$-Nash equilibrium} strategies in the influence maximization game, as defined next. 

\begin{definition}
Fix $\epsilon>0.$ A strategy profile $(\mathbf{b}_j)_{j=1}^{m} = \{(b_{1:n, 1}(k), \cdots,b_{1:n, m}(k))$, \  $k \in [K]\}$ is an $\epsilon$-Nash equilibrium if, for all influencers $j$, and for all strategies $\mathbf{b}_j'\neq \mathbf{b}_j$, we have $\frac{1}{K}\sum_{k=1}^{K} u_j(x(t_k), \rho_k,\mathbf{b}_j, \mathbf{b}_{-j})\geq \frac{1}{K}\sum_{k=1}^{K} u_j(x(t_k), \rho_k,\mathbf{b}_j', \mathbf{b}_{-j})- \epsilon$ for all $j\in[m]$. 
\end{definition}

In other words, in an $\epsilon$-Nash equilibrium, each player will be within an $\epsilon$ distance away from its best response for a given strategy of the other players, which is valid for all players. 

 \begin{corollary} \label{corr_2} 
 In the offline influence maximization game with $m\ge 2$ influencers, if every influencer $j$ plays according to a strategy with regret bound $R_j(T_r)$, then at iteration $T_r$, (i) the average strategy vector denoted by $\hat{c}_{ij}(s,\ell) =  \frac{\sum_{k=1}^{K}b_{ij}^{T_r}(k) \mathds{1}(x(k) =x^s \& T(k) = \tau_\ell)}{\sum_{k=1}^{K}\mathds{1}(x(k) =x^s \& T(k) = \tau_\ell)}$ is an $\epsilon^{T_r}$-Nash equilibrium, where $\epsilon^{T_r} = \sum_{k=1}^{T_r} \frac{R_j(T_r)}{T_r}$, and (ii) the average utility of each influencer $j$ is close to its utility at $ \hat{c}_{ij}$, that is 
 \begin{align}\nonumber
 \big|\frac{1}{T_r}\sum_{t_r=1}^{T_r}\sum_{k=1}^{K}\sum_{i=1}^{n} u_j(x(t_k),\rho_{k}, b_{j}^{t_r}(k), b_{-j}^{t_r}(k))-\sum_{k=1}^{K}\sum_{i=1}^{n} u_j(x(t_k),\rho_{k}, \hat{c}_{j}, \hat{c}_{-j})\big|\leq \sum_{j=1}^{m}\frac{R_j(T_r)}{T_r}.
 \end{align}   
 \end{corollary}

We saw earlier that the offline influence maximization game over $n$ individuals with $K$ campaign duration can be equivalently seen as a static game with $n\times K$ individuals. In Corollary~\ref{corr_2}, $T_r$ is the iteration index used in the no-regret algorithm. The proof of Corollary~\ref{corr_2} follows from the fact that in any \emph{socially concave game}, if all players follow any no-external regret procedure, then their average strategy vector converges to an $\epsilon$-Nash equilibrium of the game as shown in \cite{even2009convergence}. As we have shown in Theorem \ref{Thm_offline}, the offline influence maximization game is socially concave, and hence for $m\geq2$ we propose Algorithm~\ref{alg3:offline}, which converges efficiently to an $\epsilon$-Nash equilibrium of the offline game.  

\begin{algorithm}[t]
\begin{small}
\caption{Gradient Ascent Type Algorithm to Characterize $\epsilon$-Nash Equilibrium in the Offline Case }\label{alg3:offline}
\begin{algorithmic}[1]
\State{\textbf{Parameters:} For a given set of $\{K,\{x(t_k),T_k\}_{k\in[K]}, \{B_j\}_{j\in[m]}\}$, and Laplacian matrix $L$, average budget $\{\alpha\}_{j\in[m]}$ }
\State{\textbf{Initialize} step size $\{\eta\}_{j\in[m]}$, $\mathbf{\theta^0} = \{\theta^0_j\}_{j\in[m]}$ with $\theta^0_j>0$, $\epsilon_{thr}>0$, $\epsilon_{thr,2}>0$, \textit{cond.} is false} 
\State{\textbf{Consider} the offline static game over $n\times K$ individuals with the following modified utility functions:}
\begin{align}\label{eqn:adj_util}
    \bar{u}_j^{\ell}(\bar{x},\{\rho_k\}_{k\in[K]}, b_j, b_{-j}) = \sum_{k=1}^K\sum_{i=1}^n  \frac{\rho_{ik}(x_{ij}(k)+b_{ij}(k))}{1+\sum_{j=1}^m b_{ij}(k)} -\theta^{\ell}_j \sum_{i=1}^n b_{ij}(k)
\end{align}
\State{\textbf{while} \textit{cond.} is false}
\State{\textbf{\quad Initialize} $t_r=1$}
\State{\textbf{\quad while} $\Delta b_{t_r} > \epsilon_{thr,2}$}
\State{\textbf{\qquad Apply Online Gradient Ascent to utilities with $\bar{u}_j^{\ell}(\bar{x},\{\rho_k\}_{k\in[K]}, b_j, b_{-j})$} to find  $b_{ij}^{t_r}(k)$ }
\State{\textbf{\qquad Find $ \Delta b_{t_r}(j) = \sum_{k=1}^K\sum_{i=1}^{n}|b_{ij}^{t_r}(k)-b_{ij}^{t_r-1}(k)|$} and $\Delta b_{t_r}= \max_{j\in[m]} \Delta b_{t_r}(j)$ }
\State{\textbf{\qquad Increase $ t_r$ by 1 }}
\State{\textbf{\quad end} }
\State{\textbf{\quad Compute:} $\hat{C}_j(\ell)=\sum_{k=1}^{K}\sum_{i=1}^{n} b_{ij}^{t_r}(k)$}.
\State{\textbf{\quad Update:} $\theta_j^{\ell+1} = \theta_j^{\ell} + \eta_j (\frac{1}{Kn}\hat{C}_j(\ell) - \alpha_j)$}
\State{\textbf{\quad If $|\frac{1}{Kn}\hat{C}_j(\ell) - \alpha_j|\leq \epsilon_{thr}$ for all $j\in[m]$}}
\State{\textbf{\qquad Update:} \textit{cond.} is true}
\State{\textbf{\qquad Return:} $\tilde{b}_{ij}(k) = b_{ij}^{T_r}(k)$}
\State{\quad\textbf{end}}
\State{\textbf{end}} 
\end{algorithmic}
\end{small}
\end{algorithm}

In Algorithm~\ref{alg3:offline}, for a given set of variables, we construct the equivalent offline static game over $n\times K$ individuals with adjusted utility functions given in (\ref{eqn:adj_util}). Then, we apply an online gradient ascent-based algorithm for a given set of $\mathbf{\theta^\ell}$. At the end of iteration $T_r$, we calculate how much budget each influencer uses with $\hat{C}_j(\ell)=\sum_{k=1}^{K}\sum_{i=1}^{n} b_{ij}^{T_r}(k)$. Then, we update $\theta_j^{\ell+1} = \theta_j^{\ell} + \eta_j (\frac{1}{Kn}\hat{C}_j(\ell) - \alpha_j)$ for the next round. We repeat these steps until the average budget constraints are satisfied, i.e., $|\frac{1}{Kn}\hat{C}_j(\ell) - \alpha_j|\leq \epsilon_{thr}$ for all $j\in[m]$ for a given $\epsilon_{thr}>0$.        



\section{Nash Equilibrium Strategies for Online Multiplayer Influence Maximization Game}\label{sect:online_game}

In this section, we consider the influence maximization game with $m\geq2$ influencers and within an online setting, i.e., when the influencers cannot observe the individuals' initial opinions and the campaign durations \textit{a priori}. At the end of each round, each influencer can observe the utilities that it obtains from each individual. In particular, our goal is to show that if all influencers commit to an online strategy developed in Algorithm~\ref{alg: postcalc} under the specified information structure, then they can achieve sublinear regret compared to the Nash equilibrium in the offline game with full information about public opinions and campaign durations. Furthermore, we show that such online strategies will provide an $\epsilon$-Nash equilibrium solution, where $\epsilon$ scales with $\frac{1}{\sqrt{K}}$.\footnote{As we shall see, the information structure required for influencers to converge to their $\epsilon$-Nash equilibrium solution is slightly different for the case of $m=2$ and $m>2$ influencers.} 


\begin{theorem}\label{Theorem2}
    Let us consider the $m=2$-influencer game where the information structure of each influencer at campaign time $k$ is given by $I^{onl}_{j}(k) = \{\{x_{ij}(\ell)\}_{\ell\in[k],i\in[n]}, \{\rho_\ell\}_{\ell\in[k]},\{B_j\}_{j\in[m]}\}$. If each influencer follows Algorithm~\ref{alg: postcalc} and this is known by the other influencer, then the influencers will reach an $\epsilon$-Nash equilibrium where $\epsilon$ scales with $\frac{1}{\sqrt{K}}$.     
\end{theorem}
\begin{Proof}
First, for a given investment strategy profile of the second influencer $\mathbf{b}_2$, let us rewrite the optimization problem for the first influencer as follows:
\begin{align}\label{optimization_problem_two_player}
\max_{\{b_{i1}(k)\}} & \quad \sum_{k=1}^{K} \sum_{i=1}^n \rho_{ik}  \frac{x_{i1}(t_k) + b_{i1}(k)}{1+b_{i1}(k)+b_{i2}(k) } \nonumber\\
\textrm{s.t.} & \quad \sum_{k=1}^{K}\sum_{i=1}^{n} b_{i1}(k)\leq B_1 \nonumber\\
&\quad 0 \leq b_{i1}(k)\leq\bar{b}_1 \quad \forall i\in [n],\ \forall k \in[K].
\end{align}
Here, we note that in the online case, the future realizations of the individuals' opinions and the campaign durations are unknown. Each player can only observe the other player's strategy at the end of each round when they observe the utility obtained by each individual. The objective function in (\ref{optimization_problem_two_player}) depends on the realization of the public opinion and campaign duration, which are based on a distribution $\mathbb{P}_{\mathcal{X}\times \mathcal{T}}$, and the second influencer's investment strategy that can be arbitrarily chosen (satisfying the total budget constraint of the second influencer). Thus, the utility function at the investment opportunity $k$ given by $\sum_{i=1}^n  \rho_{ik}\frac{x_{i1}(t_k) + b_{i1}(k)}{1+b_{i1}(k)+b_{i2}(k)}$ can be considered to be chosen adversarially that is unknown to the first influencer. In this case, we can assume that the utility functions are obtained based on a distribution with infinite support.\footnote{As noted in \cite{balseiro2020dual}, Algorithm~\ref{alg: postcalc} will have the same regret bound in the case of infinite support.} As in the single influencer case, all the assumptions of \cite{balseiro2020dual} also hold for the online optimization problem  (\ref{optimization_problem_two_player}). Thus, by using Theorem~\ref{Thm_1}, for a fixed strategy of the second influencer, if the first influencer applies Algorithm~\ref{alg: postcalc}, its regret bound scales with $\sqrt{K}$. Thus, for the first influencer, we have 
\begin{align}
    \sum_{k=1}^{K} u_1(x(t_k), \rho_k,\mathbf{\hat{b}}_1, \mathbf{b}_{2})\geq \sum_{k=1}^{K} u_1(x(t_k), \rho_k,BR_1(\mathbf{b}_{2}), \mathbf{b}_{2})- \kappa \sqrt{K},
\end{align}
where $BR_1(\mathbf{b}_{2})$ denotes the best response of influencer $1$ for a given other influencer's strategy $\mathbf{b}_{2}$, $\mathbf{\hat{b}_1}$ denotes the policy as a result of following Algorithm~\ref{alg: postcalc}, and $\kappa>0$ is a finite constant. Since this inequality holds for $BR_1(\mathbf{b}_{2})$, it also holds for any other arbitrarily selected policy such that $\mathbf{b}_1^{'} \neq \mathbf{\hat{b}}_1$, and we have  
\begin{align}\label{thm2_eqn1}
    \sum_{k=1}^{K} u_1(x(t_k), \rho_k,\mathbf{\hat{b}}_1, \mathbf{b}_{2})\geq \sum_{k=1}^{K} u_1(x(t_k), \rho_k,\mathbf{b}_1^{'}, \mathbf{b}_{2})- \kappa \sqrt{K}.
\end{align}
Similarly, for a given investment strategy of the first influencer $ \mathbf{b}_1$, let $\mathbf{\hat{b}}_2$ be the strategy of the second influencer obtained by following Algorithm~\ref{alg: postcalc}. Then, we have 
\begin{align}\label{thm2_eqn2}
    \sum_{k=1}^{K} u_1(x(t_k), \rho_k,\mathbf{b}_{1}, \mathbf{\hat{b}}_2)\geq \sum_{k=1}^{K} u_1(x(t_k), \rho_k,\mathbf{b}_{1}, \mathbf{b}_{2}^{'})- \kappa^{'}\sqrt{K},
\end{align}
where $\kappa^{'} >0$. Thus, if we denote the joint online strategy of the influencers obtained by following Algorithm~\ref{alg: postcalc} by $(\mathbf{\hat{b}}_1,\mathbf{\hat{b}}_2)$, we will have 
\begin{align} \label{thm2_eqn3}
    \frac{1}{K}\sum_{k=1}^{K} u_{1}(x(t_k), \rho_k,\mathbf{\hat{b}}_{1}, \mathbf{\hat{b}}_2)&\geq \frac{1}{K}\sum_{k=1}^{K} u_1(x(t_k), \rho_k,\mathbf{b}_{1}', \mathbf{\hat{b}}_{2})- \epsilon,\nonumber\\ 
    \frac{1}{K}\sum_{k=1}^{K} u_{2}(x(t_k), \rho_k,\mathbf{\hat{b}}_{1}, \mathbf{\hat{b}}_2)&\geq \frac{1}{K}\sum_{k=1}^{K} u_2(x(t_k), \rho_k,\mathbf{\hat{b}}_{1}, \mathbf{b}_{2}^{'})- \epsilon,
\end{align}
where $ \epsilon = \mathcal{O}(\frac{1}{\sqrt{K}})$. This shows that if both influencers follow Algorithm~\ref{alg: postcalc}, they will reach an $\epsilon$-Nash equilibrium of the offline game, where $\epsilon=\mathcal{O}(\frac{1}{\sqrt{K}})$. 
\end{Proof}

Theorem~\ref{Theorem2} shows that if both the influencers follow the same online algorithm, they will reach an $\epsilon$-Nash equilibrium. In the proof of Theorem~\ref{Theorem2}, equations (\ref{thm2_eqn1}) and (\ref{thm2_eqn2}) are obtained based on the fact that influencer $j$ applies the online algorithm based on the given investment strategy of the other influencer. However, in the $\epsilon$-Nash equilibrium equations given in (\ref{thm2_eqn3}), we inherently require influencer $j$ to know the investment strategy of the influencer $-j$. This is possible in the following way. We assume that each influencer can observe the individuals' opinions $x(t_k)$ and the campaign duration at the beginning of each round, and also they observe the utilities obtained by each influencer at the end of each round, which is given by $\frac{\rho_{ik}(x_{ij}(k) + b_{i1}(k))}{1+ b_{i1}(k)+ b_{i2}(k)}$. From the utilities obtained at round $k$, the influencer $j$ can find $ b_{i,-j}(k)$. Since both influencers apply Algorithm~\ref{alg: postcalc}, with the information available at step $k$, they can infer the investment strategy of the other influencer at round $k+1$, i.e., $b_{i,-j}(k+1)$. Thus, we may assume that $b_{i,-j}(k+1)$ is known by the influencer $j$ at the end of round $k$, and the $\epsilon$-Nash equilibrium in (\ref{thm2_eqn3}) is well-defined.

\begin{corollary}\label{corr_asymp}
In the two-player online influencer game with the information structure $I^{onl}_{j}(k)$, the average utility obtained at an $\epsilon$-Nash equilibrium by applying Algorithm~\ref{alg: postcalc} asymptotically equals the average  utility obtained by the unique Nash equilibrium strategy in the offline game.  
\end{corollary}
\begin{Proof}
    Let us denote the unique Nash equilibrium in the offline game by the pair $(\mathbf{b_{1}^*},\mathbf{b_{2}^*})$. From the definition of Nash equilibrium, we have
\begin{align} \label{eqn_nash_corr}
    \sum_{k=1}^{K} u_{1}(x(t_k), \rho_k,\mathbf{b_{1}^*}, \mathbf{b_{2}^*})&\geq \sum_{k=1}^{K} u_1(x(t_k), \rho_k,\mathbf{b}_{1}', \mathbf{b_{2}^*}), \quad \forall \mathbf{b}_{1}',\nonumber\\ 
    \sum_{k=1}^{K} u_{2}(x(t_k), \rho_k,\mathbf{b_{1}^*}, \mathbf{b_{2}^*})&\geq \sum_{k=1}^{K} u_2(x(t_k), \rho_k,\mathbf{b_{1}^*}, \mathbf{b}_{2}^{'}), \quad \forall \mathbf{b}_{2}'.
\end{align}
Then, by using (\ref{thm2_eqn3}) and (\ref{eqn_nash_corr}), we obtain:
\begin{align}
 \frac{1}{K}\sum_{k=1}^{K} u_{2}(x(t_k), \rho_k,\mathbf{\hat{b}}_{1}, \mathbf{\hat{b}}_2)&\geq \frac{1}{K}\sum_{k=1}^{K} u_2(x(t_k), \rho_k,\mathbf{\hat{b}}_{1}, \mathbf{b}_{2}^{*})- \epsilon, \label{eqn_temp}\\
 \frac{1}{K}\sum_{k=1}^{K} u_{1}(x(t_k), \rho_k,\mathbf{b_{1}^*}, \mathbf{b_{2}^*})&\geq \frac{1}{K}\sum_{k=1}^{K} u_1(x(t_k), \rho_k,\mathbf{\hat{b}}_{1}, \mathbf{b_{2}^*}).\label{eqn_temp_2}
\end{align}
Since the game is a constant-sum game, i.e., $\sum_{k=1}^{K} \big(u_{1}(x(t_k), \rho_k,\mathbf{b}_{1}, \mathbf{b}_2)+u_{2}(x(t_k), \rho_k,\mathbf{b}_{1}, \mathbf{b}_2)\big) = \sum_{k=1}^{K}\rho_k$, for any pair $\{\mathbf{b}_{1}, \mathbf{b}_2\}$, if we subtract $\sum_{k=1}^{K}\rho_k$ in (\ref{eqn_temp}) from both sides, we obtain
\begin{align}
\frac{1}{K}\sum_{k=1}^{K} u_{1}(x(t_k), \rho_k,\mathbf{\hat{b}}_{1}, \mathbf{\hat{b}}_2)-\epsilon\leq \frac{1}{K}\sum_{k=1}^{K} u_1(x(t_k), \rho_k,\mathbf{\hat{b}}_{1}, \mathbf{b}_{2}^{*}).\label{eqn_temp_3}
\end{align}
By combining (\ref{eqn_temp_2}) and (\ref{eqn_temp_3}), we have:
\begin{align}
 \frac{1}{K}\sum_{k=1}^{K} u_{1}(x(t_k), \rho_k,\mathbf{\hat{b}}_{1}, \mathbf{\hat{b}}_2)-\epsilon\leq \frac{1}{K}\sum_{k=1}^{K} u_{1}(x(t_k), \rho_k,\mathbf{b_{1}^*}, \mathbf{b_{2}^*}). \label{eqn_temp_4} 
\end{align}
Using a similar argument, we can write 
\begin{align}
 \frac{1}{K}\sum_{k=1}^{K} u_{2}(x(t_k), \rho_k,\mathbf{\hat{b}}_{1}, \mathbf{\hat{b}}_2)-\epsilon\leq\frac{1}{K} \sum_{k=1}^{K} u_{2}(x(t_k), \rho_k,\mathbf{b_{1}^*}, \mathbf{b_{2}^*}).\label{eqn_temp_5}  
\end{align}
If we subtract $\sum_{k=1}^{K}\rho_k$ from (\ref{eqn_temp_5}), we get
\begin{align}
 \frac{1}{K}\sum_{k=1}^{K} u_{1}(x(t_k), \rho_k,\mathbf{\hat{b}}_{1}, \mathbf{\hat{b}}_2)+\epsilon\geq \frac{1}{K}\sum_{k=1}^{K} u_{1}(x(t_k), \rho_k,\mathbf{b_{1}^*}, \mathbf{b_{2}^*}).\label{eqn_temp_6}  
\end{align}
Finally, by using (\ref{eqn_temp_6}) and (\ref{eqn_temp_4}), we have
\begin{align}
 \frac{1}{K}\sum_{k=1}^{K} u_{1}(x(t_k), \rho_k,\mathbf{\hat{b}}_{1}, \mathbf{\hat{b}}_2)-\epsilon\leq \frac{1}{K}\sum_{k=1}^{K} u_{1}(x(t_k), \rho_k,\mathbf{b_{1}^*}, \mathbf{b_{2}^*})\leq\frac{1}{K} \sum_{k=1}^{K} u_{1}(x(t_k), \rho_k,\mathbf{\hat{b}}_{1}, \mathbf{\hat{b}}_2)+\epsilon. \label{eqn_temp_7} 
\end{align}
Since $\epsilon$ converges to $0$ at a rate of $\mathcal{O}(\frac{1}{\sqrt{K}})$, the average utility obtained by the $\epsilon$-Nash equilibrium strategy in the online case is the same as the utility obtained by the unique Nash equilibrium in the offline case. 
\end{Proof}

Finally, we turn our attention to the online influence maximization game with $m>2$ influencers. The following theorem provides a performance guarantee for the online Algorithm~\ref{alg: postcalc} once followed by all the influencers in the game.  

\begin{theorem}\label{Theorem3}
    Consider the online influencer game with $m>2$ influencers, where the information structure of each player $j$ is given by $\hat{I}^{onl}_{j}(k) = \{\{x_{ij}(\ell)\}_{i\in[n],j\in[m],\ell\in[k]}, \{\rho_\ell\}_{\ell\in[k]},\{B_j\}_{j\in[m]}, \boldsymbol{\theta^0}\}$ for all $j\in[m]$. If all influencers commit to the online policy in Algorithm~\ref{alg: postcalc} with initial $\boldsymbol{\theta^0}= \{\theta^0_1,\theta^0_2,\cdots,\theta^0_m\}$, then the influencers will reach an $\epsilon$-Nash equilibrium with $\epsilon=\mathcal{O}(\frac{1}{\sqrt{K}})$.     
\end{theorem}
\begin{Proof}
The proof follows similar steps as in the proof of Theorem~\ref{Theorem2}. The main difference is that in the case of $m=2$-influencer game, by observing the utilities obtained from each individual, influencer $j$ can infer influencer $-j$'s investment strategy. On the other hand, in the case of $m>2$-influencers, by observing the utilities obtained from each individual, influencer $j$ knows $\sum_{\ell\neq j}b_{i\ell}(k)$, which is not enough to determine $b_{ij}(k)$ at campaign time $k$.       

For given strategies of other influencers $-j$, influencer $j$ wants to maximize its own utility given by 
\begin{align}\label{optimization_problem_m_player}
\max_{\{b_{ij}(k)\}} & \quad \sum_{k=1}^{K} \sum_{i=1}^n \rho_{ik}  \frac{x_{ij}(t_k) + b_{ij}(k)}{1+\sum_{j=1}^m b_{ij}(k) } \nonumber\\
\textrm{s.t.} & \quad \sum_{k=1}^{K}\sum_{i=1}^{n} b_{ij}(k)\leq B_j \nonumber\\
&\quad 0 \leq b_{ij}(k)\leq\bar{b}_j, \quad \forall i\in [n],\ \forall k \in[K].
\end{align}
If all influencers apply Algorithm~\ref{alg: postcalc}, then at campaign opportunity $k$, influencer $j$ finds its resource allocation based on the following equation:
\begin{align}\label{eqn:opt_online_m_players}
  \tilde{b}_{ij}(k) = \min &\left\{ \left( \sqrt{\frac{\rho_{ik}(1-x_{ij}(k)+ \sum_{j`\neq j} \tilde{b}_{ij`}(k))}{\theta_j^k}}-1-\sum_{j`\neq j} \tilde{b}_{ij`}(k)\right)^+, \bar{b}_j\right\}. 
\end{align}
We note that in order to find the online allocation strategy at stage $k$, influencer $j$ needs to anticipate the allocation strategy of the others at stage $k$, i.e., $\tilde{b}_{ij`}(k)$ for all $ j`\neq j$. Each influencer commits to Algorithm~\ref{alg: postcalc} and knows the opinion realizations for all influencers, i.e., $\{x_{ij}(k)\}_{i\in[n],j\in[m]}$, initial budgets $B_j$ for $j\in[m]$, and initial values of $\boldsymbol{\theta}^0$. Therefore, under such an information structure, each influencer can form (\ref{eqn:opt_online_m_players}) for all $j$, and thus they can solve for $\tilde{b}_{ij}(k)$ for all $i\in[n]$ and $j\in[m]$. Finally, each influencer obtains its investment strategy by using (\ref{eqn:opt_onl_investment}).

Therefore, under the information structure $\hat{I}^{onl}_{j}(k)$, if all influencers follow Algorithm~\ref{alg: postcalc}, denoted by their joint online strategy $(\mathbf{\hat{b}}_1,\cdots,\mathbf{\hat{b}}_m)$, we will have 
\begin{align} \label{thm3_eqn3}
    \frac{1}{K}\sum_{k=1}^{K} u_{j}(x(t_k), \rho_k,\mathbf{\hat{b}}_{j}, \mathbf{\hat{b}}_{-j})&\geq \frac{1}{K}\sum_{k=1}^{K} u_j(x(t_k), \rho_k,\mathbf{b}_{j}', \mathbf{\hat{b}}_{-j})- \epsilon,
\end{align}
where $\epsilon = \mathcal{O}(\frac{1}{\sqrt{K}})$ and $\mathbf{b}_{j}'$ is any arbitrary policy such that  $\mathbf{b}_{j}'\neq \mathbf{\hat{b}}_{j}$. Since (\ref{thm3_eqn3}) holds for all $j\in[m]$, we conclude that the influencers will reach an $\epsilon$-Nash equilibrium, where $\epsilon$ scales with $\frac{1}{\sqrt{K}}$.     
\end{Proof}

We note that in the case of $m>2$ influencer game, in order to reach an $\epsilon$-Nash equilibrium solution, influencers' information structure needs to be $\hat{I}^{onl}_{j}(k)$, which may not be feasible in some scenarios. For that reason, in the following, we propose an alternative online algorithm that relaxes this assumption and requires less information structure. Then, we provide a condition under which the alternative online algorithm performs the same as Algorithm~\ref{alg: postcalc} under the richer information structure $\hat{I}^{onl}_{j}(k)$. 

To that end, by assuming that $0\leq \tilde{b}_{ij}(k)\leq \bar{b}_j$, we can rewrite (\ref{eqn:opt_online_m_players}) as  
\begin{align}\label{eqn:opt_online_m_players_v2}
\tilde{b}_{ij}(k) = - \frac{\theta_j^k}{\rho_{ik}}\tilde{G}_{ij}(k)^2+\tilde{G}_{ij}(k)-x_{ij}(k),
\end{align}
where $\tilde{G}_{ij}(k) = 1 +\sum_{j=1}^{m} \tilde{b}_{ij}(k)$. Using some algebraic manipulations, we find $\tilde{G}_{ij}(k) = \frac{(m-1)\rho_{ik}}{\sum_{j=1}^m \theta_j^k}$. Thus, (\ref{eqn:opt_online_m_players_v2}) becomes 
\begin{align}\label{eqn:opt_online_m_players_v3}
\tilde{b}_{ij}(k) = \frac{(m-1)\rho_{ik}}{\sum_{j=1}^m \theta_j^k}\left(1- \frac{(m-1)\theta_j^k}{\sum_{j=1}^m \theta_j^k}\right)- x_{ij}(k).
\end{align}
The advantage of (\ref{eqn:opt_online_m_players_v3}) is that in order to find $\tilde{b}_{ij}(k)$ values, the influencers only need to know $\boldsymbol{\theta^k}$, $\boldsymbol{\rho_{k}}$, and $x_{ij}(k)$, which requires less information as compared to $\hat{I}^{onl}_{j}(k)$. In fact, such an information structure can be easily implemented in practical situations using a central authority that reports $\boldsymbol{\theta^{k}}$ vector to all influencers at the beginning of each campaign time $k$. Then, each influencer applies $\tilde{b}_{ij}(k)$ in (\ref{eqn:opt_online_m_players_v3}) and reports their remaining budget $B_j(k+1)$ to the central authority at the end of the campaign time. Based on the remaining budgets, the central authority updates $\boldsymbol{\theta^{k}}$ vector and sends it back to influencers for campaign time $k+1$. 

We note that the above derivations are valid under the assumption that $0\leq \tilde{b}_{ij}(k)\leq \bar{b}_j \forall j$. Therefore, in the following corollary, we provide some conditions under which this assumption is always satisfied.   


\begin{corollary}
    Let us consider the case where $\bar{b}_j = \bar{b}$ for all $j$ and assume that $\bar{b}$ is large enough so that $\tilde{b}_{ij}(k)\leq \bar{b}$ is always satisfied. Then, if the following conditions hold 
    \begin{align}\label{eqn:corr_final_v2}
         \theta_{j}^k\leq \frac{\rho_{ik}}{m\bar{b}}\left( 1-\frac{1}{m\bar{b}}\right), \qquad \theta_{j}^{max} = \frac{\bar{u}}{\alpha_j}+1 \leq \frac{\rho_{ik}}{m\bar{b}}\left( 1-\frac{1}{m\bar{b}}\right),
    \end{align}
   then we have $0\leq\tilde{b}_{ij}(k)$.
\end{corollary}
\begin{Proof}
    First, we note from (\ref{eqn:opt_online_m_players_v3}) that if we have 
\begin{align} \label{eqn:corr_1}
   1\leq \frac{(m-1)\rho_{ik}}{\sum_{j=1}^m \theta_j^k}\left(1- \frac{(m-1)\theta_j^k}{\sum_{j=1}^m \theta_j^k}\right)\leq \bar{b}, 
\end{align}
then $0\leq\tilde{b}_{ij}(k)\leq \bar{b}$. By using (\ref{eqn:corr_1}), we obtain $\frac{(m-1)\rho_{ik}}{m\bar{b}}\leq \sum_{j=1}^{m} \theta_{j}^k\leq \frac{(m-1)\rho_{ik}}{m}$. As we assume that $\bar{b}$ is large enough so that the upper bound $\tilde{b}_{ij}(k)\leq \bar{b}$ is always satisfied, we consider the lower bound in (\ref{eqn:corr_1}), i.e., $1\leq \frac{(m-1)\rho_{ik}}{\sum_{j=1}^m \theta_j^k}\left(1- \frac{(m-1)\theta_j^k}{\sum_{j=1}^m \theta_j^k}\right)$, which implies that $\theta_j^k\leq \frac{\sum_{j=1}^m \theta_j^k}{m-1}\left( 1-\frac{\sum_{j=1}^m \theta_j^k}{(m-1)\rho_{ik}}\right) $. Since $\sum_{j=1}^{m} \theta_{j}^k\in[\frac{(m-1)\rho_{ik}}{m\bar{b}},\frac{(m-1)\rho_{ik}}{m} ]$, the tightest upper bound on $\theta_j^k$ can be found as 
\begin{align}
    \theta_j^k\leq \frac{\rho_{ik}}{m\bar{b}}\left( 1-\frac{1}{m\bar{b}}\right).
\end{align}
Finally, in Algorithm~\ref{alg: postcalc}, we defined $\theta^{max}_j = \frac{\bar{u}}{\alpha_j}+1$, where $\bar{u} = \sum_{i=1}^{n}\rho_{ik}$, which was used to derive the regret bound in Section~\ref{Sect:single}. If $\theta^{max}_j\leq \frac{\rho_{ik}}{m\bar{b}}\left( 1-\frac{1}{m\bar{b}}\right)$ is satisfied for all $j$, $\theta^{max}_j$ will lie within the bound, and thus the regret analysis in Section~\ref{Sect:single} would follow. 
\end{Proof}

Under the assumption that $\bar{b}$ is large enough, the first condition in (\ref{eqn:corr_final_v2}) ensures that $0\leq\tilde{b}_{ij}(k)$, which can enable influencers to have less information and they are still able to find $\epsilon$-Nash equilibrium strategies with the help of a central authority. 
As we will show in our numerical results,  
the above alternative online algorithm with less information closely mimics the behavior of the online Algorithm \ref{alg: postcalc} with the richer information structure $\hat{I}^{onl}_{j}(k)$, and serves as a very good approximation to it.

\section{Numerical Results}\label{Sect:num_res}

In this section, we provide numerical results to illustrate the theoretical analysis found in the previous sections. In the first set of simulations, we consider a single influencer and compare the online and offline strategies. In this example, there are $n=4$ individuals, and each individual's opinion takes an independent realization from $0.1$ to $0.9$. Thus, $\mathcal{X} = \{x^r|x^r\in [0,1]^{1\times4}, r= 1,\dots, 9^4\}$ with $\mathbb{P}_{\mathcal{X}}(x^r) = \frac{1}{9^4}$ for all $r$. The campaign duration $T_k$ takes independent values from $1$ to $10$ uniformly. Thus, we have $\mathcal{T} = \{1,\dots, 10\}$ with pmf given by $\mathbb{P}_{\mathcal{T}}(\tau) = 0.1$. The individuals in the network interact with each other based on the De-Groot model ($\dot{x}(t) = -Lx(t)$ for $t_{k-1}\leq t\leq  t_k$), where $L$ is given by
\begin{align}\label{L_matrix}
 L =\begin{bmatrix}
3 &  -1&  -1 & -1\\
-\frac{1}{3}& 2  & -\frac{2}{3}&  -1\\
0 &-1 &1   & 0\\
-1& -2 &  0   &3
\end{bmatrix}.
\end{align}

\begin{figure}[t]	\centerline{\includegraphics[width=0.60\columnwidth]{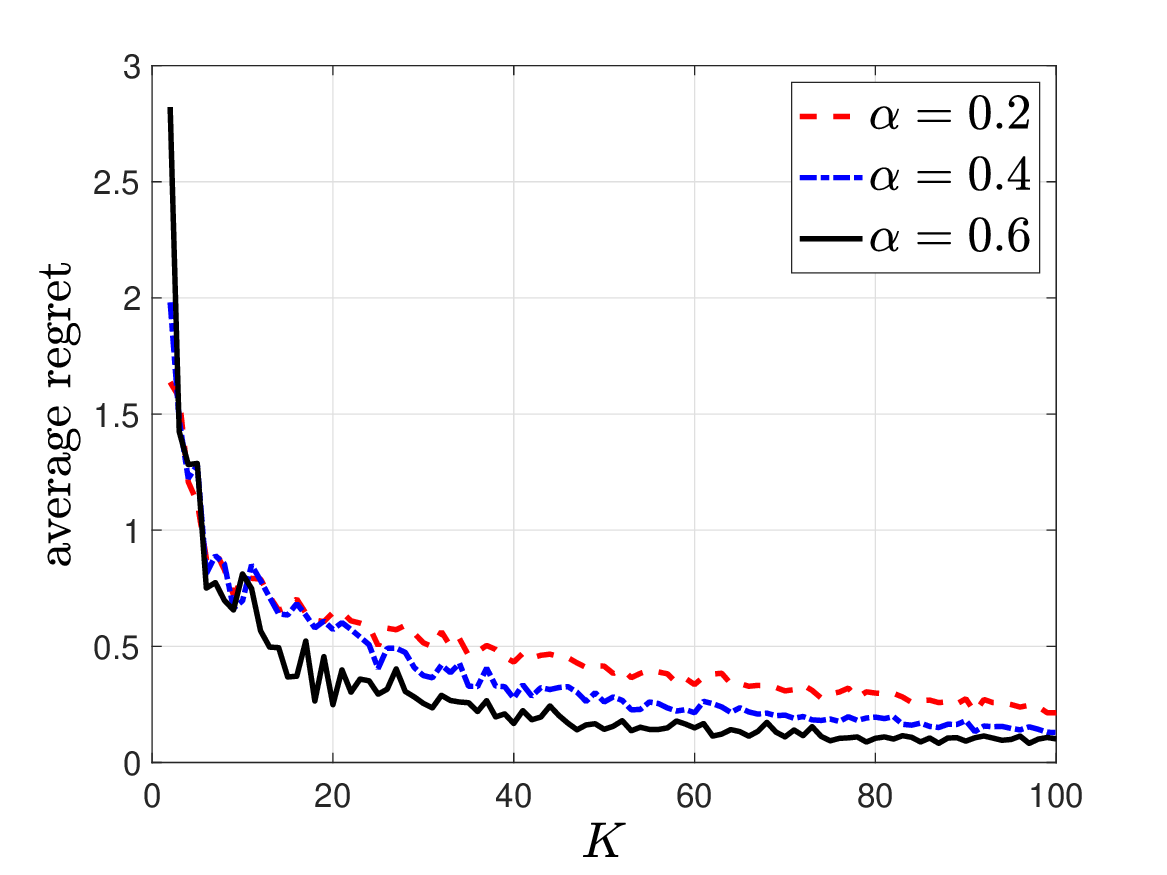}}
	\caption{Average regret for a single influencer when $\alpha \in\{ 0.2, 0.4, 0.6\}$.}
\label{Fig:Inf_flow}
\vspace{-0.5cm}
\end{figure}

In this numerical example, we increase the number of campaign opportunities $K$ from 2 to 100 and consider three different average budgets for the influencer, which are given by $\alpha = \{0.2, 0.4, 0.6\}$, where $\alpha = \frac{B}{K}$. We compare the performances of the online and offline strategies by looking at the average regret of the influencer. In Fig.~\ref{Fig:Inf_flow}, we see that for all $\alpha$ values, the average regret converges to 0, meaning that the average utilities obtained by the online and offline strategies converge as $K$ grows. We also observe in Fig.~\ref{Fig:Inf_flow} that when the influencer has a limited budget, i.e., when $\alpha$ is low, the convergence is slower. This is reasonable because as the budget gets lower, the influencer must carefully allocate its budget to individuals. In the online case in which the influencer does not get to observe the opinions' realizations \textit{a priori}, it will take a longer time to learn the optimal allocation strategy.    

\begin{figure}[t]
	\begin{center}
	    \subfigure[]{%
			\includegraphics[scale=0.4]{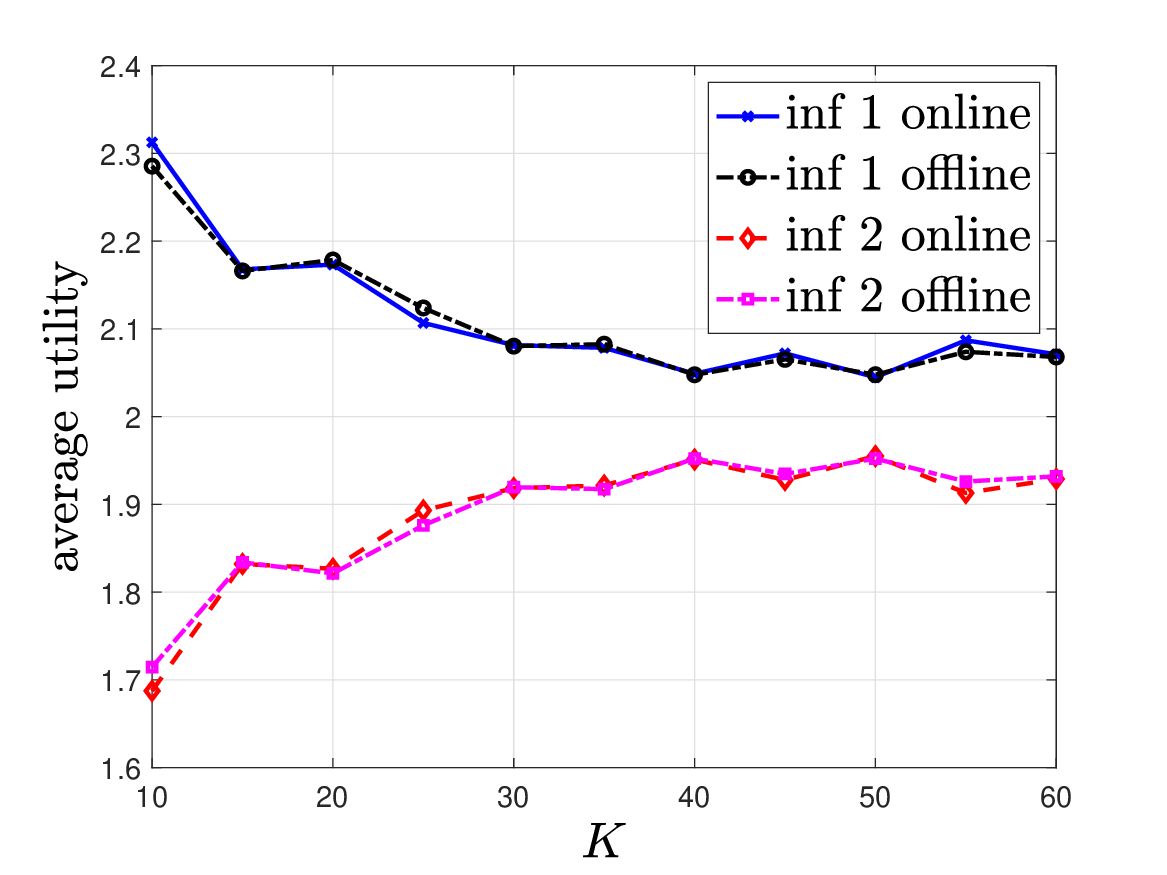}}
		\subfigure[]{%
			\includegraphics[scale=0.4]{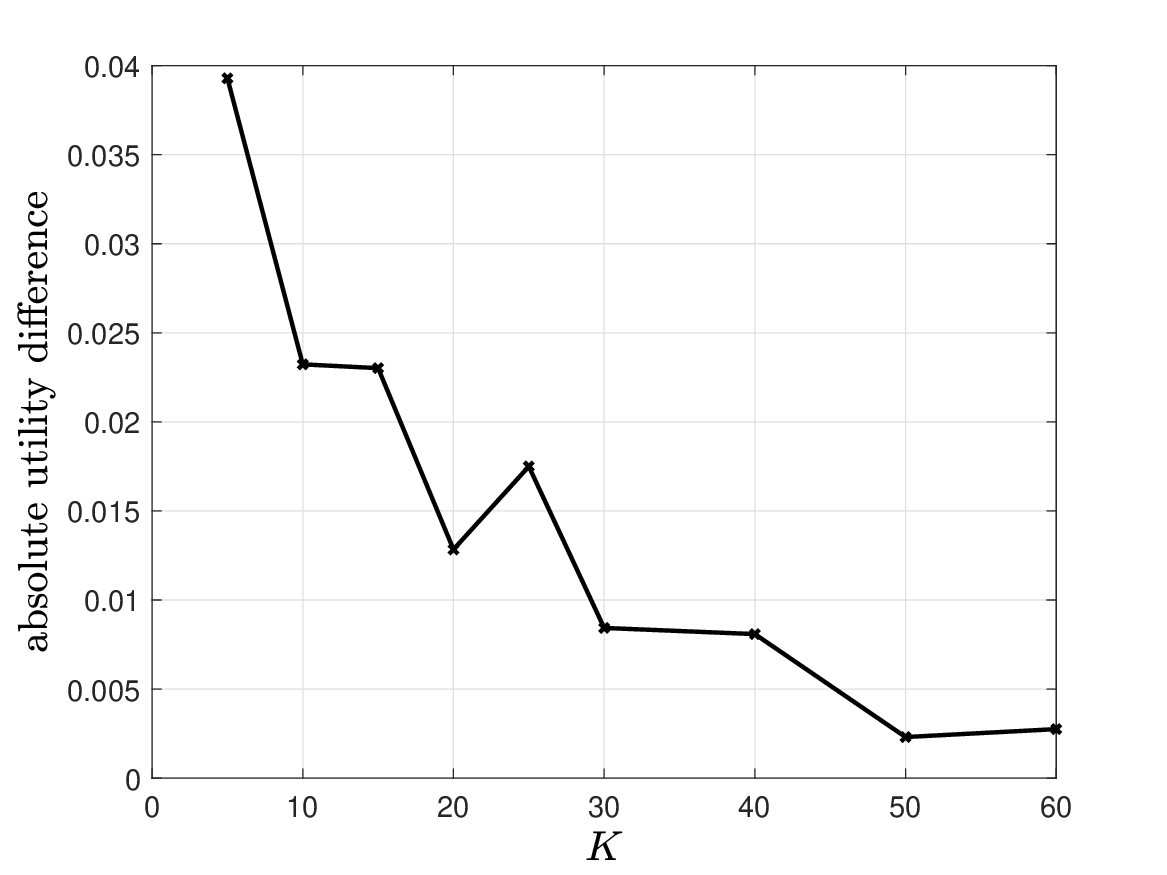}}
	\end{center}\vspace{-0.5cm}
	\caption{Average utility difference of the online and offline strategies in the presence of $m=2$ influencers.}
	\label{fig:sim2-sec}
	\vspace{-0.5cm}
\end{figure}

In our second numerical result, we consider $m=2$ influencers with $\alpha_1 = 0.7$ and $\alpha_2 = 0.5$, i.e., the first influencer has a higher budget compared to the second influencer. We consider that each individual's opinion for the first influencer can take $x_{i1}(t_k)\in \{0.2,0.5,0.8\}$ values uniformly at random. Similarly, opinion dynamics for the second influencer is $x_{i2}(t_k) = 1-x_{i1}(t_k) \in \{0.2,0.5,0.8\}$. We note that this particular opinion dynamics can represent the scenario where the individual favors, is neutral, or opposes an influencer. The campaign duration can take values $T_k\in \{2,4,6,8\}$ uniformly at random. We take $n=4$, and the Laplacian matrix $L$ is the same as in the previous example. In this numerical result, we compare the performance of the offline and online strategies as the number of campaign opportunities $K$ increases from 10 to 60. In the offline case, we showed in Theorem~\ref{Thm_offline} that there is a unique Nash equilibrium among the influencers. For the $m=2$ influencer game, we find the unique Nash equilibrium solution by applying the best response dynamics repeatedly. For the online case, each influencer follows Algorithm~\ref{alg: postcalc}, they can reach $\epsilon$-Nash equilibrium solutions, where $\epsilon=\mathcal{O}(\frac{1}{\sqrt{K}})$ as shown in Theorem~\ref{Theorem2}. We provide the utilities obtained from each influencer in Fig.~\ref{fig:sim2-sec}(a). As the first influencer has a higher budget, the first influencer gets more utility compared to the second one. The average utilities obtained in the online and offline policies are close to each other. More precisely, we plot $\frac{1}{K}|\sum_{k=1}^{K} u_{1}(x(t_k), \rho_k,\mathbf{\hat{b}_{1}}, \mathbf{\hat{b}_2})-  \sum_{k=1}^{K} u_{1}(x(t_k), \rho_k,\mathbf{b_{1}^*}, \mathbf{b_{2}^*})|$ in Fig~\ref{fig:sim2-sec}(b), where $(\mathbf{b_{1}^*}, \mathbf{b_{2}^*})$ denotes the Nash equilibrium strategy and $(\mathbf{\hat{b}}_{1}, \mathbf{\hat{b}}_2)$ denotes the $\epsilon$-Nash equilibrium strategy obtained by following Algorithm~\ref{alg: postcalc}. We observe in Fig~\ref{fig:sim2-sec}(b) that the average utility difference between the online and offline strategies converges to zero as $K$ increases.          

\begin{figure}[t]
	\begin{center}
	    \subfigure[]{%
			\includegraphics[scale=0.4]{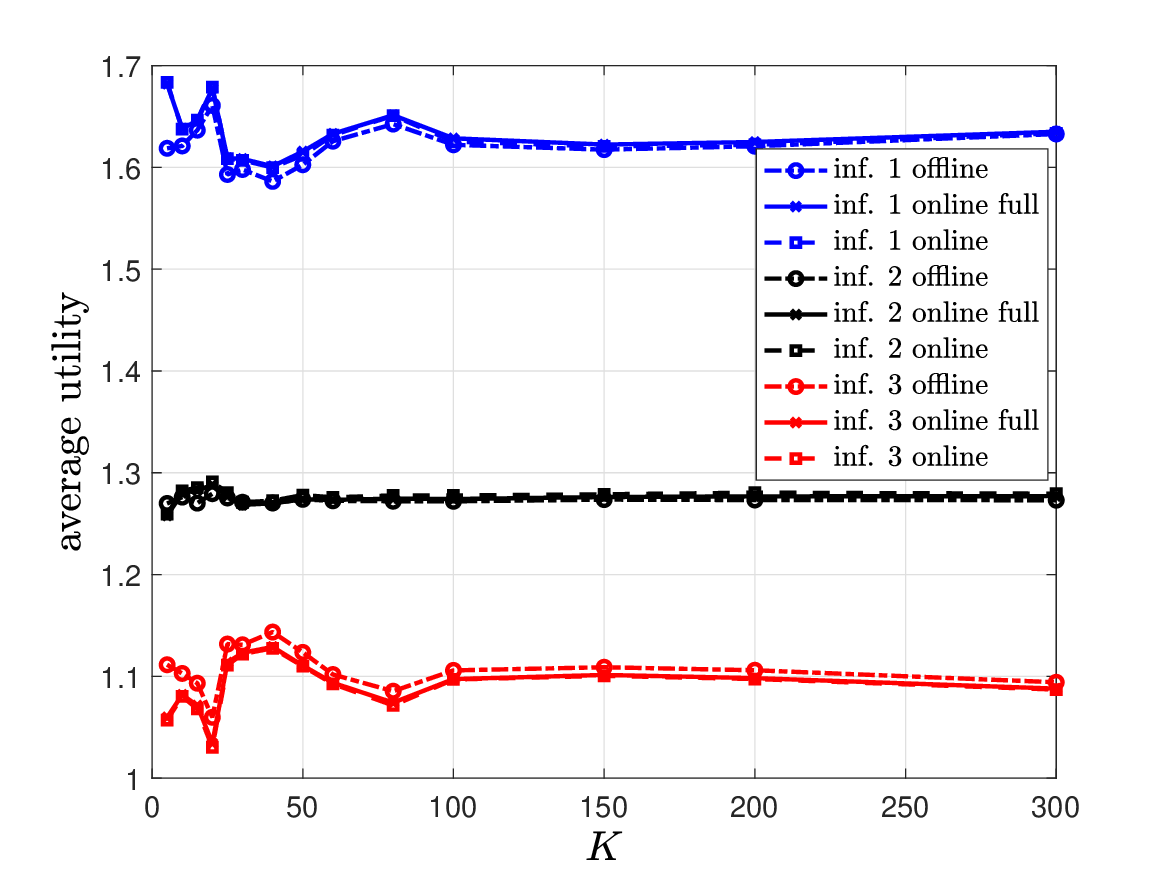}}
		\subfigure[]{%
			\includegraphics[scale=0.4]{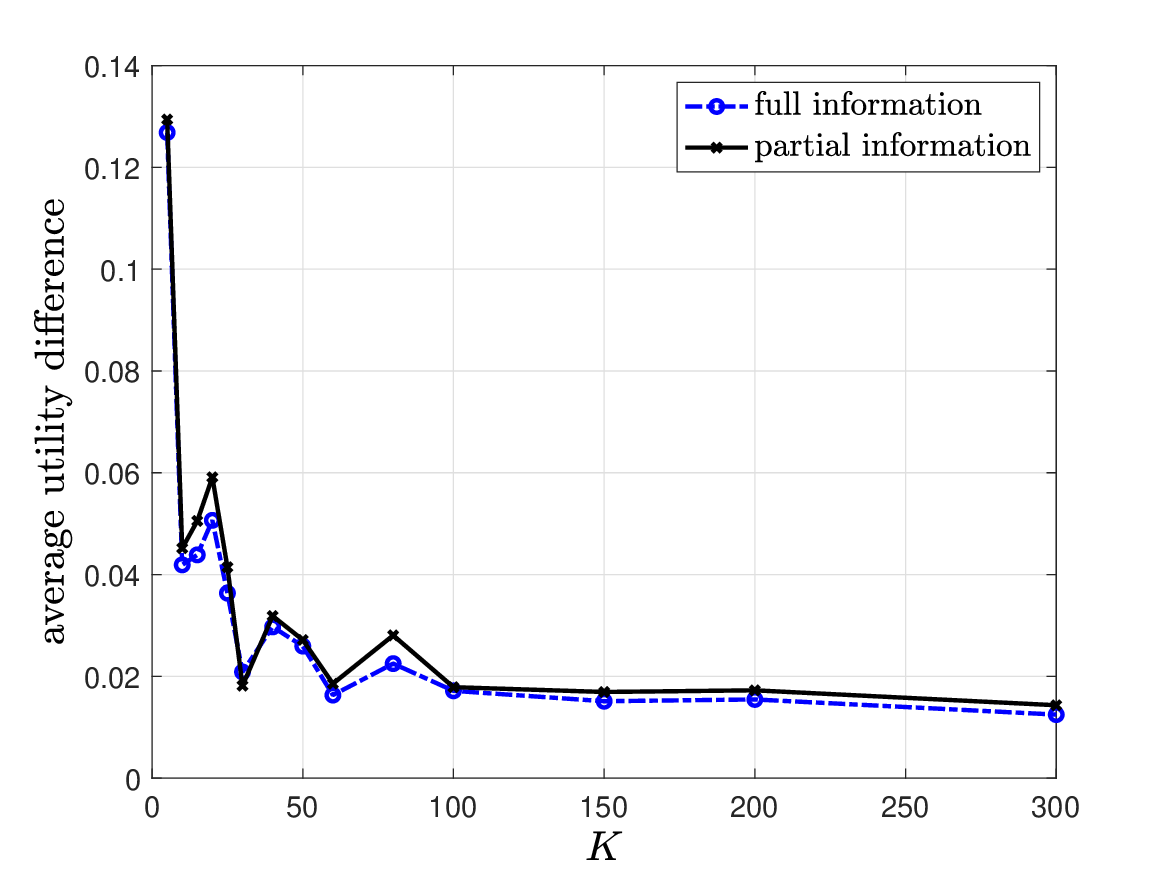}}
	\end{center}\vspace{-0.5cm}
	\caption{(a) The average utilities of the 3 influencer system as a result of a result of applying offline, online with full information, and online with partial information. (b) The total average utility difference of the online policies compared to the offline policy.}
	\label{fig:sim3}
	\vspace{-0.5cm}
\end{figure}

In the third numerical example, we consider $m=3$ influencer system with $\alpha_1 = 0.7$, $\alpha_2= 0.5$, and $\alpha_3 = 0.3$. Here, each individual's opinion takes independent realizations from the set $x_{i,1:3}\in\{ [0.2,0.3,0.5], [\frac{1}{3},\frac{1}{3},\frac{1}{3} ],[0.6,0.3,0.1]\}$ uniformly. Note that with these realizations, an individual can be initially in favor of influencer 1 and against influencer 3 (when $x_{i,1:3}= [0.2,0.3,0.5]$), neutral to all influencers (when $x_{i,1:3}= [\frac{1}{3},\frac{1}{3},\frac{1}{3} ]$), or in favor of influencer 3 and against influencer 1 (when $x_{i,1:3}= [0.6,0.3,0.1]$). We consider $n=4$, and the $L$ matrix is the same as in the previous example. In this numerical example, since we consider more than two influencers, as mentioned earlier, the best response dynamics may not return the Nash equilibrium solution. However, as shown in Corollary~\ref{corr_2} for the offline case, if every influencer $j$ plays a no-regret algorithm, they will reach an $\epsilon$-Nash equilibrium. For that, we propose Algorithm~\ref{alg3:offline}, which utilizes online gradient ascent to find the allocation policy. For the online case, we consider two settings for the information structure: \textit{full information setting} given by $\hat{I}^{onl}_{j}(k) = \{\{x_{ij}(\ell)\}_{i\in[n],j\in[m],\ell\in[k]}, \{\rho_\ell\}_{\ell\in[k]},\{B_j\}_{j\in[m]}, \boldsymbol{\theta^0}\}$ for all $j\in[m]$ and \textit{partial information setting} with $\tilde{I}^{onl}_{j}(k) = \{\{x_{ij}(\ell)\}_{i\in[n],\ell\in[k]}, \{\rho_\ell\}_{\ell\in[k]},B_j, \theta^k_j, \sum_{j=1}^{m}\theta^k_j\}$. In other words, with the full information setting, influencers know the entire realization of the opinion dynamics and initial $\boldsymbol{\theta^0}$, and thus can follow the remaining budgets of the other influencers. With the full information setting, starting with (\ref{eqn:opt_online_m_players_v3}) as the initial solution, influencers run (\ref{eqn:opt_online_m_players}) until the allocation policy converges. In the case of oscillating between different allocation policies, we choose any one of them arbitrarily. Considering the scenarios where full information setting is not feasible, we propose a way to find an allocation policy with partial information setting $\tilde{I}^{onl}_{j}(k)$, which requires each influencer to know its own opinion realization, its own remaining budget, and $\sum_{j=1}^{m}\theta^k_j$. Then, the influencers can find their allocation policy by using (\ref{eqn:opt_online_m_players_v3}). \footnote{In order to obtain $\sum_{j=1}^{m}\theta^k_j$, the influencer may utilize a central agent which collects $\theta_j^k$ from the influencers and reports $\sum_{j=1}^{m}\theta^k_j$ truthfully.} \footnote{In other words, both the full information and partial information settings are using (\ref{eqn:opt_online_m_players_v3}) as their initial solution. However, with the full information setting, the influencers can further iterate their initial solution with using (\ref{eqn:opt_online_m_players}). } We see the average utilities obtained by the influencers in Fig.~\ref{fig:sim3}(a). Online policies with full and partial information settings perform closely with the offline setting. In Fig.~\ref{fig:sim3}(b), we plot $\frac{1}{K}\sum_{j=1}^{m}|\sum_{k=1}^{K} u_{j}(x(t_k), \rho_k,\mathbf{\hat{b}_{1}}, \mathbf{\hat{b}_2},\mathbf{\hat{b}_3})-  \sum_{k=1}^{K} u_{j}(x(t_k), \rho_k,\mathbf{b_{1}^*}, \mathbf{b_{2}^*},\mathbf{b_{3}^*})|$. We observe that as $K$ increases, the online strategy obtained as a result of having full information performs better than the online strategy with partial information. However, we see that online strategy with partial information performs well and could be preferred as working with $\hat{I}^{onl}_{j}(k)$ may not always be possible. 

\begin{figure}[t]
	\begin{center}
 		
	    \subfigure[]{%
			\includegraphics[scale=0.27]{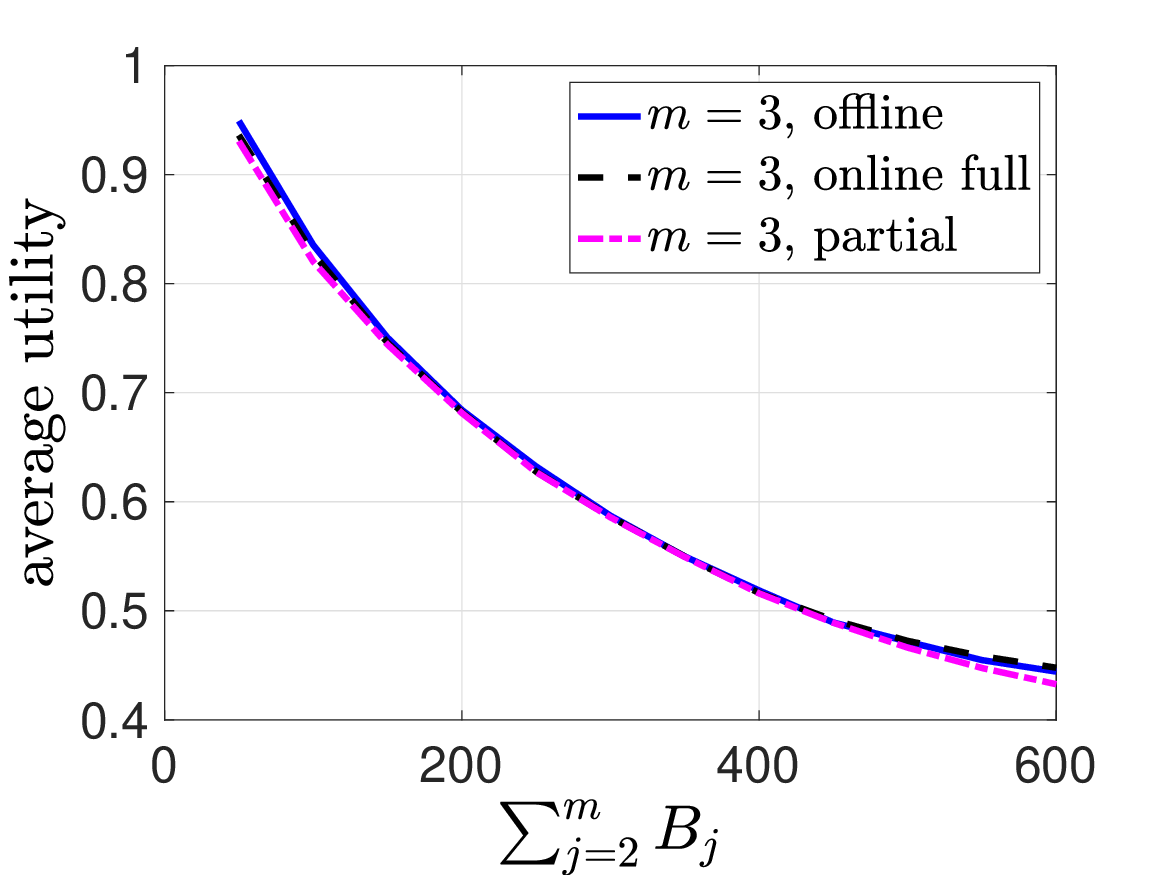}}
		\subfigure[]{%
			\includegraphics[scale=0.27]{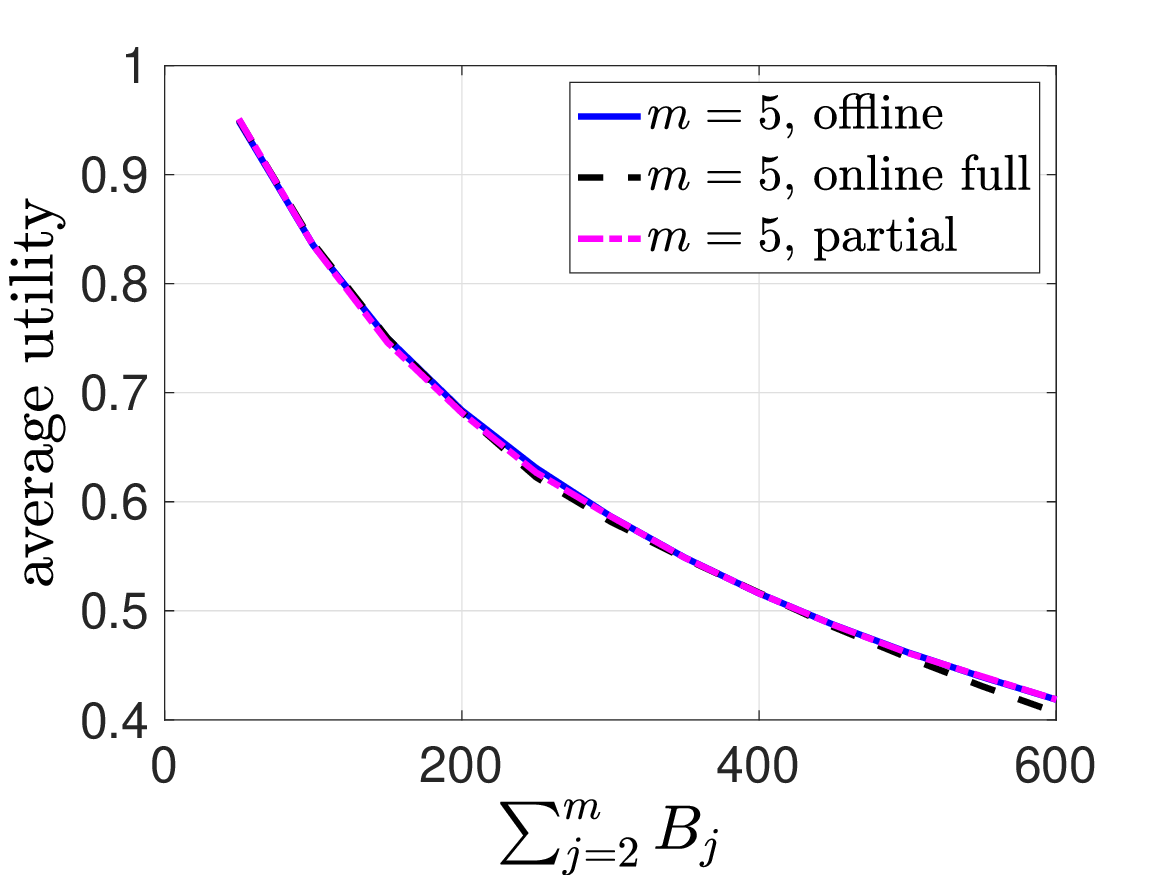}}
   	    \subfigure[]{%
			\includegraphics[scale=0.27]{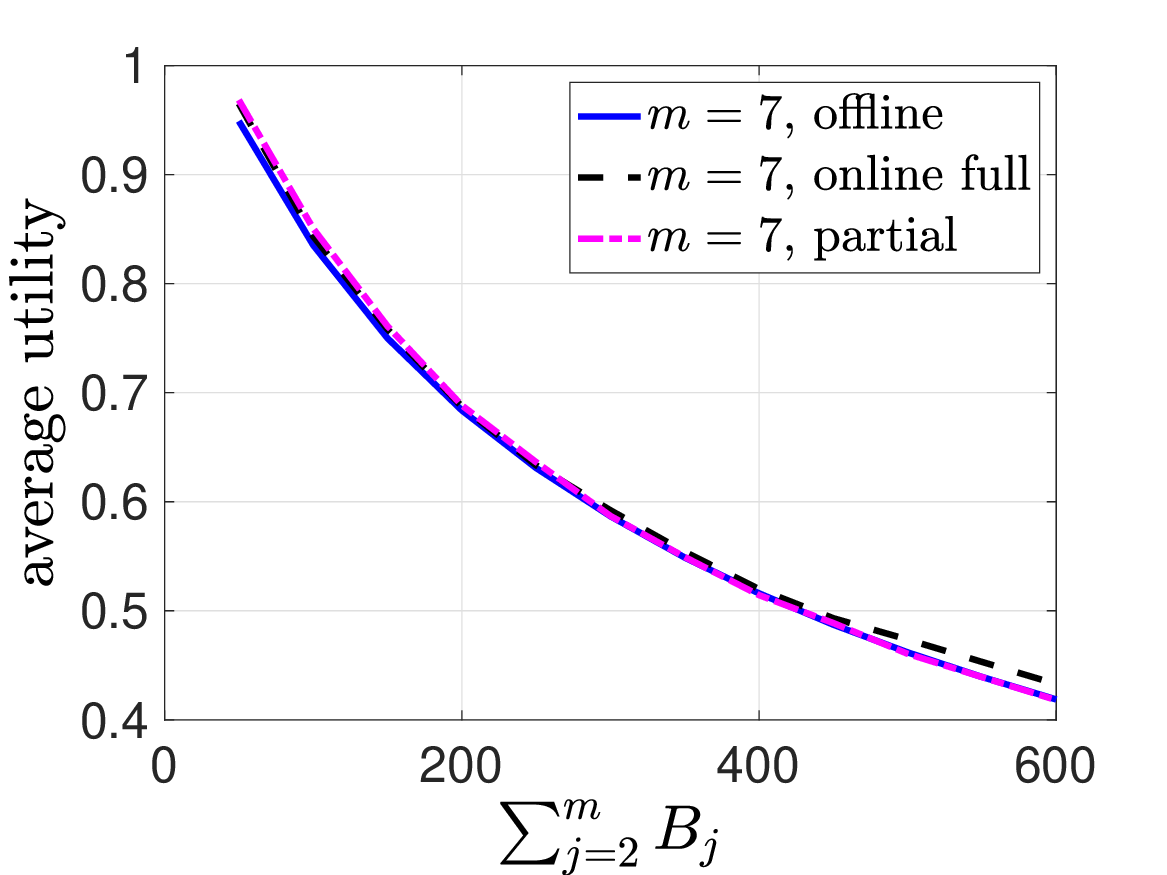}}\\
		\subfigure[]{%
			\includegraphics[scale=0.27]{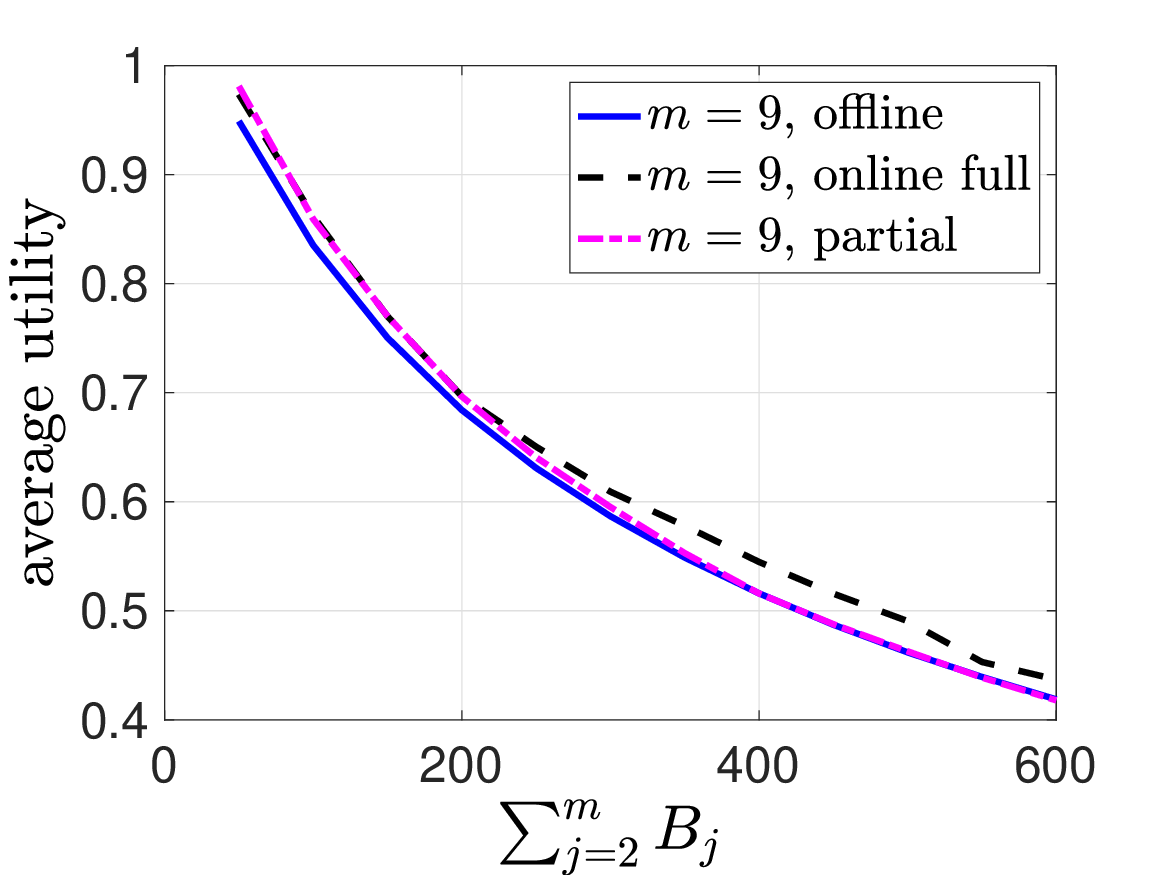}}
   \subfigure[]{%
			\includegraphics[scale=0.27]{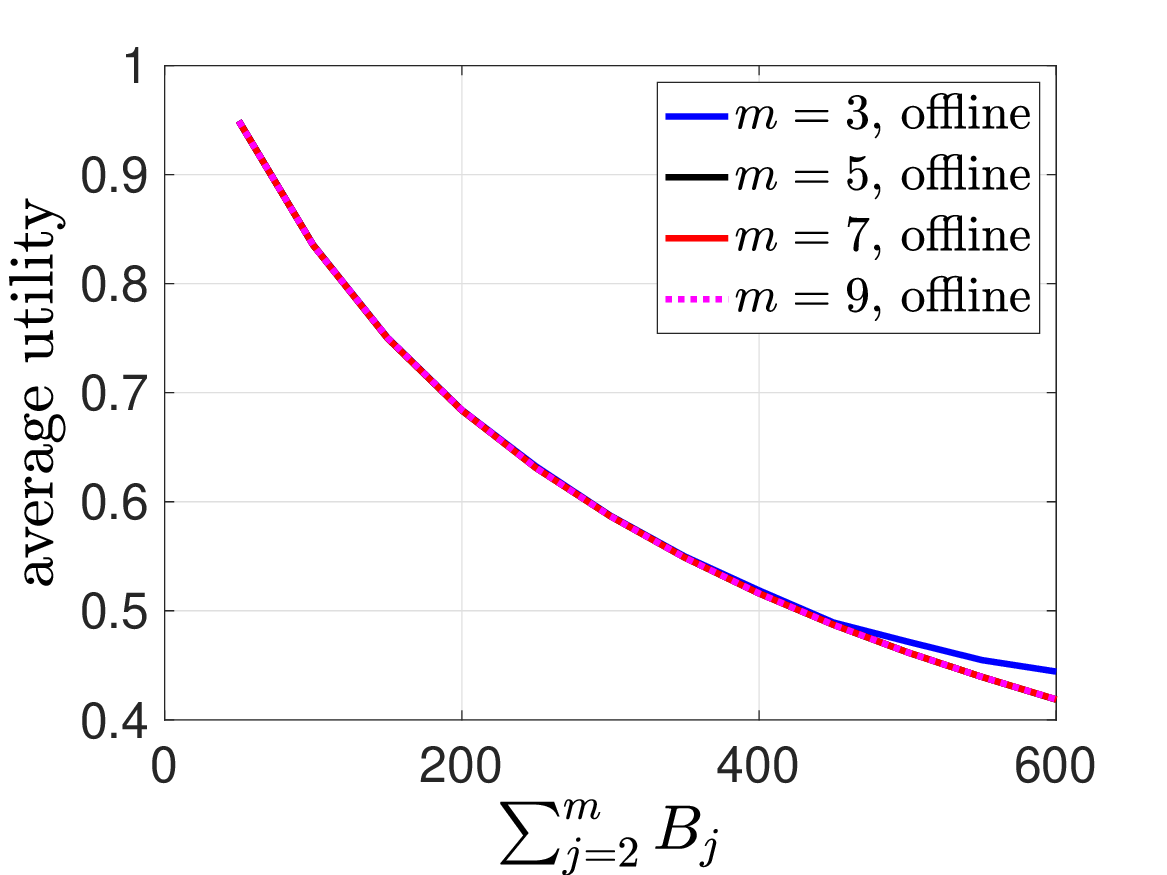}}
      \subfigure[]{%
			\includegraphics[scale=0.27]{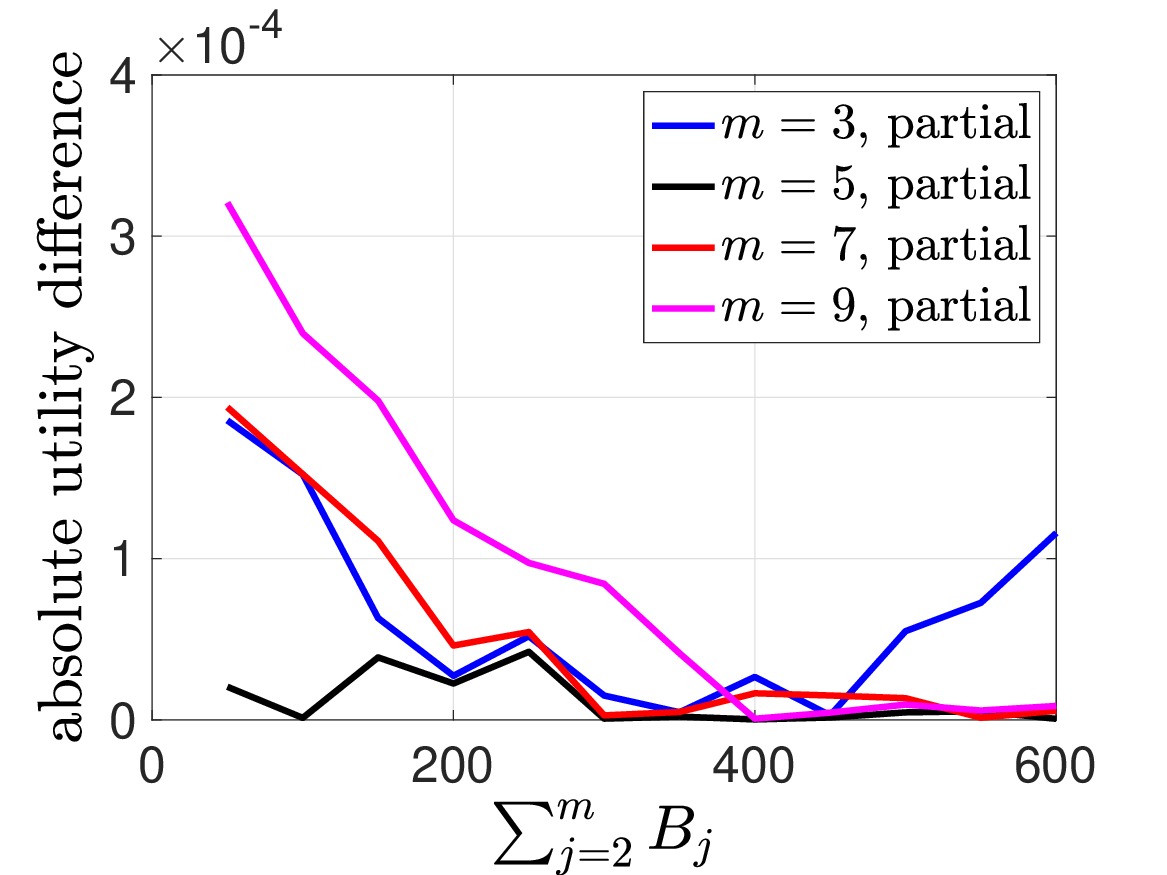}}
	\end{center}\vspace{-0.5cm}
	\caption{The average utility obtained from influencer 1 while increasing the sum of the other influencers budget when (a) $m=3$, (b) $m=5$, (c) $m=7$, and (d) $m=9$. (e) The comparison of the average utility of influencer 1 as a result of applying the offline policy when $m \in \{3,5,7,9\}$.}
	\label{fig:sim6}
	\vspace{-0.5cm}
\end{figure}

In the fourth numerical example, we consider the utility of the first influencer as we increase the total budget of the other influencers when the total numbers of influencers are $m=3,5,7,9$. We take $n=4$, $K=100$, and the $L$ matrix is the same as in the previous 2 examples, $\bar{b} = 1$, $B_1 = 50$ (and thus, $\alpha_1 = \frac{B_1}{K} = 0.5$). The opinion dynamics of the individuals toward the first influencer take values from the set $x_{i,1}(k)= \{0.1, 0.2, 0.3, 0.4\}$ uniformly. We consider the initial opinion dynamics for the other influencers as $x_{i,j}(k) = \frac{1-x_{i,1}(k)}{m-1}$ for $j=2,\dots,m$. For instance, if $x_{i,1}(k) = 0.4$ and $m=3$, then we have $x_{i,2}(k)$ = $x_{i,3}(k) = 0.3$. The campaign times are equal to $T_k = 5$ for all $k\in[K]$. In this numerical example, while we keep the first influencer's budget $B_1$ the same, we increase the total budget of the other influencers, i.e., $\sum_{j=2}^{m} B_j$ from 50 to 600 where they each have the same total budget, i.e., $B_2= \dots = B_m$. In Figs.~\ref{fig:sim6}(a)-(b)-(c)-(d), we provide the utility of the first influencer obtained from the offline policy, online policies with full and partial information when $m=3,5,7,9$, respectively. Here, we see that as we increase the total budget of the other influencers, the utility of the first influencer decreases, which is aligned with intuition. In all cases, the online policies perform closely to the offline policies. Interestingly, as we increase $m$, we see in Figs.~\ref{fig:sim6}(c)-(d) that the online policy with partial information performs better than the online policy with full information which may indicate that our proposed solution may not benefit from the full information setting fully. Then, we compare the utility of the first influencer for the offline case, which is the utility of the first influencer obtained as a result of the $\epsilon$-Nash solution as we increase $m$ from 3 to 9 in Fig.~\ref{fig:sim6}(e). Here, at least with the settings considered in this example, we see that the utility of the first influencer is similar as we increase the number of influencers but we keep the total budget of the other influencers at the same level. Finally, we plot $\frac{1}{K}|\sum_{k=1}^{K} u_{1}(x(t_k), \rho_k,\mathbf{\hat{b}_{1}}, \mathbf{\hat{b}_{-1}})-  \sum_{k=1}^{K} u_{j}(x(t_k), \rho_k,\mathbf{b_{1}^*}, \mathbf{b_{-1}^*})|$ in Fig.~\ref{fig:sim6}(f) which shows that the online policy with partial information performs closely with the offline policies.

\section{Conclusion and Future Directions}
In this work, we have studied dynamic influence maximization games for multiple influencers who have total budget rate constraints. First, we considered the influence maximization problem for a single influencer, where we provided solutions for the offline and online cases. In particular, Algorithm~\ref{alg: postcalc} has a sublinear regret rate. Then, we focused on game formulations with multiple influencers. For the offline case, we showed that the dynamic game over $K$ campaign times with $n$ individuals can be considered a static game over $n\times K$ individuals with a unique Nash equilibrium due to social concavity property. As for the online case, when we have $m=2$ influencers and each influencer follows Algorithm~\ref{alg: postcalc}, we showed that the influencers converge to an $\epsilon$-Nash equilibrium. We extended this result to the general number of influencers under a more strict information structure. We also developed an alternative online algorithm with less information, which can be run independently by the influencers to approximate the offline $\epsilon$-Nash equilibrium strategies. We have seen in our fourth set of numerical results in Fig.~\ref{fig:sim6} that our proposed online policy with partial information performs as good as the one with the full information setting. As an immediate extension to our work, we would like to further investigate proposing another online algorithm that can benefit from the full information setting fully. 

As a future direction, one can also consider the influence maximization game in which the opinion realizations are not independent at the beginning of the campaign opportunities but rather are correlated based on a Markov chain model, leading to a Markov game formulation. Also, one can consider a direction where the opinion dynamics evolve continuously in the form of a differential game between the campaign times. In this case, influencers need to plan their investment strategies carefully, as their current investments would also affect the evolution of future opinion dynamics.

\bibliographystyle{unsrt}
\bibliography{IEEEabrv,myLibrary_bastopcu}

\end{document}